\documentclass{aa}  

\usepackage{graphicx}
\usepackage{subcaption}
\usepackage{txfonts}
\usepackage{hyperref}
\usepackage{comment}
\usepackage{lineno}

\hypersetup{
    colorlinks=true,                
    breaklinks=true,                
    urlcolor= red,                  
    linkcolor= blue,                
    citecolor= blue                 
    }

\defcitealias{Plavin_2019}{P19}

\begin{document} 

   \title{Dynamic and Radiative Implications of Jet-Star Interactions in AGN Jets}

   \titlerunning{Dynamic and Radiative Implications of Jet-Star Interactions in AGN Jets}

   \author{G. Fichet de Clairfontaine
          \inst{1},
          M. Perucho
          \inst{1, 2},
          J. M. Martí
          \inst{1, 2}
          \and 
          Y. Y. Kovalev
          \inst{3}
          }

    \authorrunning{Fichet de Clairfontaine et al.}

   \institute{Departament d’Astronomia i Astrof\'isica, Universitat de Val\`encia, C/ Dr. Moliner, 50, E-46100 Burjassot, Val\`encia, Spain\\
              \email{gaetan.fichet@uv.es}
         \and
             Observatori Astron\`omic, Universitat de Val\`encia, C/ Catedr\`atic Jos\`e Beltr\`an 2, E-46980 Paterna, Val\`encia, Spain 
         \and 
            Max-Planck-Institut für Radioastronomie, Auf dem Hügel 69, 53121 Bonn, Germany
             }

   \date{Accepted for publication in A\&A.}

\abstract
{The interactions between jets from active galactic nuclei (AGN) and their stellar environments significantly influence jet dynamics and emission characteristics. In low-power jets, such as those in Fanaroff-Riley I (FR I) galaxies, the jet-star interactions can notably affect jet deceleration and energy dissipation.}
{Recent numerical studies suggest that mass loading from stellar winds is a key factor in decelerating jets, accounting for many observed characteristics in FR I jets. Additionally, a radio-optical positional offset has been observed, with optical emission detected further down the jet than radio emission. This observation may challenge traditional explanations based solely on recollimation shocks and instabilities.}
{This work utilizes the radiative transfer code \texttt{RIPTIDE} to generate synthetic synchrotron maps, from a population of re-accelerated electrons, in both radio and optical bands from jet simulations incorporating various mass-loading profiles and distributions of gas and stars within the ambient medium.}
{Our findings emphasize the importance of mass entrainment in replicating the extended and diffuse radio/optical emissions observed in FR\,I jets and explaining the radio-optical offsets. These offsets are influenced by the galaxy's physical properties, the surrounding stellar populations, and observational biases. We successfully reproduce typical radio-optical offsets by considering a mass-load equivalent to $10^{-9}~M_\odot \cdot \rm{yr}^{-1}\cdot{pc}^{-3}$. Overall, our results demonstrate that positive offset measurements are a promising tool for revealing the fundamental properties of galaxies and potentially their stellar populations, particularly in the context of FR I jets.}
{}

   \maketitle
%

\section{Introduction}
\label{sec:Introduction}

Jets from active galactic nuclei (AGN) represent some of the most powerful and persistent structures in the universe. Given the complexity and simultaneous occurrence of high-energy processes within these jets, an obvious first step was to categorize them. According to the Fanaroff-Riley (FR) classification \citep{Fanaroff_1974}, radio-loud AGN are divided into FR\,I and FR\,II types based on large-scale morphology and radio luminosity. We observe key distinctions between FR\,I and FR\,II jets across two primary dimensions: dynamics and emission characteristics. Dynamically, FR\,I jets become decollimated at kiloparsec scales and appear to decelerate, whereas FR\,II jets remain collimated and retain relativistic speeds over larger distances. In terms of emission, FR\,I jets exhibit diffuse and extended radio emission, while FR\,II jets display localized emission at knots within the jet or at termination lobes 
\citep[e.g.,][]{Briddle_1984}. The reasons behind these differences remain under investigation. It is uncertain whether the observed distinctions are either solely due to variations in processes near the jet base—such as during jet launching or acceleration phases—or whether they extend to phenomena occurring along the jet propagation \citep{Laing_1996}. Nevertheless, observations at kiloparsec scales suggest that both jet types maintain relativistic speeds, underscoring the significant role of the surrounding ambient medium at these scales in influencing the deceleration and morphological characteristics of FR\,I jets \citep{Laing_2014}.

Interactions between FR\,I jets and their environments can manifest in various ways. A frequently considered mechanism involves the entrainment of colder, denser ambient gas during jet propagation \citep{Bicknell_1984, Bicknell_1994}. Non-linear perturbations such as pinching, triggered by recollimation shocks \citep{Falle_1991,  Perucho_2007, Mizuno_2015, Fichet_2021, Fichet_2022}, or relativistic centrifugal instabilities \citep{Matsumoto_2013, Matsumoto_2017, Gourgoliatos_2018} can destabilize the jet, facilitating local matter entrainment and contributing to its deceleration. Despite extensive observations, definitive conclusions regarding these phenomena remain elusive \citep{perucho_2019}.

An intriguing hypothesis is that jet deceleration occurs through mass loading from stellar mass loss, a scenario proposed by \cite{Komissarov_1994} and extensively explored through numerical simulations \citep{Bowman_1996, Perucho_2014, Angles_2021}. Jet-star interactions are inevitable, with their occurrences potentially reaching up to $10^8$ within the first kiloparsec of the jet \citep{Vieyro_2017}. This process is treated as a hydrodynamical problem, with mass-loading conceptualized as a mass injection along the jet trajectory. Simulations indicate that mass-loading in low-power jets ($L_{\rm j} = 10^{42-43}~\rm{erg}\cdot\rm{s}^{-1}$) causes deceleration, potentially enhancing jet-interstellar medium mixing, which can further slow the jet. If the jet initially comprises electron-positron pairs, mixing with the material can transform it into a lepto-hadronic (electron-proton) jet, increasing local internal energy through dissipation and potentially leading to the production of non-thermal emissions \citep{Perucho_2017, Vieyro_2017, Torres_2019}. Additionally, a toroidal magnetic field configuration may limit jet expansion and thereby influence mass-loading effects \citep{Angles_2021}.

The hydrodynamical implications of mass-loading are critical for understanding jet dynamics, especially regarding the complexity and diversity of the physics at play \citep{Boccardi_2017, Blandford_2019}. The impact of mass-loading could be linked to the presence of radio-optical offsets. These offsets have been observed and described intensively in the past \citep[e.g.,][]{Petrov_2017l, Kovalev_2017, Petrov_2017,Petrov_2019, Plavin_2019,Kovalev_2020,Plavin_2022,Secrest_2022, Lambert_2024}. Particularly, \cite{Plavin_2019} (below shown as \citetalias{Plavin_2019}) analyzed the variation of peak emission positions between sources detected using very-long-baseline interferometry (VLBI) in the radio spectrum and \emph{Gaia} data release~2 (DR\,2) in the optical spectrum. They quantified the offset between the radio to optical peak coordinate positions and the jet flow direction through the angle $\Psi$ (see their Figure~1). When $\Psi \sim 0^\circ$, the \emph{Gaia} position is located downstream from the VLBI position, indicating a positive radio-optical offset. Conversely, $\Psi \sim 180^\circ$ indicates that the \emph{Gaia} peak position is closer to the jet base compared to the radio peak, indicating a negative radio-optical offset.

After cross-identifying a sample of $4023$ sources, including quasars, BL Lacertae (BL~Lacs), Seyfert $1$ and $2$, radio galaxies, and unknown types, \citetalias{Plavin_2019} observed anisotropy in the $\Psi$ distribution, with two dominant values: $\Psi \sim 0^\circ$ and $\Psi \sim 180^\circ$. This distribution suggests that the offset is significantly influenced by the contribution of the jet to the total optical emission. They found that jets with $\Psi \sim 0^\circ$ constitute the majority of AGN with significant detectable offset, exhibiting a median radio-optical offset ranging from $0.70$ milliarcseconds (mas) for quasars to $7.2~\rm{mas}$ for Seyfert $2$ galaxies.
For the population where $\Psi \sim 180^\circ$, the influence of the accretion disk is suggested, with bluer \emph{Gaia} index colors, thus moving the optical centroid towards the AGN nucleus. For positive radio-optical offsets ($\Psi \sim 0^\circ$), a bright and extended optical jet could account for the positional discrepancy. The study concluded that the sign of the offset (positive or negative) reflects a competition between the contributions from the accretion disk (negative offset) and the jet (positive offset). Radio-optical offsets characteristics for various types of AGN mentioned are displayed in Table~\ref{tab:plavin_results}. These findings have been confirmed in a more recent study by \cite{Lambert_2024}.

\renewcommand{\arraystretch}{1.2}
\begin{table*}
    \caption{Object type and their median radio-optical offset value, sign (positive or negative), and \emph{Gaia} color index (evaluated in the blue and red bands) as written in \citetalias{Plavin_2019}.}
    \label{tab:plavin_results}
    \centering
    \begin{tabular}{lcccc}
        \hline\hline
        AGN type & $\Psi~\left[^\circ\right]$ & Median offset~$\left[\rm{mas}\right]$ & Sign offset~$\left[+/-\right]$ & Typical \emph{Gaia} color index~$\left[\rm{B}-\rm{R}\right]$ \\
        \hline 
        Quasar & $0^\circ/180^\circ$ & $0.7$ & $+/-$ & $0.6/0.6$ \\
        BL Lac & $0^\circ$ & $0.62$ & $+$ & $1.2$ \\
        Seyfert $1$ & $0^\circ/180^\circ$ & $0.73$ & $+/-$ & $1.5/0.5$ \\
        Seyfert $2$ & $0^\circ$ & $7.2$ & $+$ & $1.5$ \\
        Radio galaxies & $0^\circ$ & $2.3$ & $+$ & ? \\
        \hline
    \end{tabular}
\end{table*}
\renewcommand{\arraystretch}{1}

Considering only diffusive acceleration and synchrotron radiation from a population of non-thermal electrons, such an offset implies a distinct dissipation process occurring further from the base of the jet, leading to optical emissions spatially separated from those in the radio spectrum. This process likely results in increased internal energy of the fluid, thereby elevating the minimum Lorentz factor, $\gamma^\prime_{\rm e, min}$, of the non-thermal electrons. Essentially, the internal energy input raises the average kinetic energy of the electrons, which shifts the entire energy distribution upwards, including the minimum Lorentz factor of the non-thermal population. In this scenario, the synchrotron radiative cooling is expected to be negligible, and thus not taken into account. The gain of energy through increase of internal energy is expected to be the dominant effect here. As the mass-loading scenario shows promising results in these objects, one should try to reproduce the presence of such radio-optical offsets through it. Notably, the star mass-loading hypothesis elucidates the rise in internal energy at a specific distance from the jet base \citep{Bowman_1996, Perucho_2014, Angles_2021}.

Motivated by these observations, this paper aims to delve into the radiative contributions from the mass-loaded jets previously detailed in \cite{Angles_2021}. We explore various mass-loading profiles to elucidate the presence of positive radio-optical offsets and their characteristics. Employing the RIPTIDE code \citep{Fichet_2021, Fichet_2022}, we compute synthetic synchrotron maps that consider relativistic effects and the source distance from Earth. Our analysis of these maps provides insights into the nature of the observed offsets, enhancing our understanding of jet dynamics and interactions within their host galaxies. Moreover, our results open the possibility of deriving information about host galaxies using jets as a probe.

The paper is organized as follows. Sect.~\ref{sec:Numerical setup} describes our numerical setup, which consists of the numerical code used to carry the jet simulations and the radiative transfer code. Then, Sect.~\ref{sec:Results} presents our results. We discuss these results in light of the past radio-optical offset observations and other observational evidence in Sect.~\ref{sec:Discussion}, before presenting our conclusions in Sect.~\ref{sec:Conclusions}. Throughout this paper, quantities given in the jet co-moving frame are primed, and we assume a flat $\Lambda$CDM cosmology with $\rm{H}_{\rm 0} = 69.6~\rm{km}\cdot\rm{s}^{-1}\cdot\rm{Mpc}^{-1}$, $\Omega_{\rm 0} = 0.29$ and $\Omega_{\Lambda} = 0.71$. 


\section{Numerical setup}
\label{sec:Numerical setup}

\subsection{Jet simulation}
\label{subsec:Jet simulation}

As presented in \cite{Angles_2021} the simulations have been done following the approach described in \citep{Komissarov_2015}. According to it, under the approximations of a narrow jet (jet radius much smaller than its length) and a flow speed close to the speed of light, models of steady axially-symmetric jets can be built by solving the time-dependent (magneto-) hydrodynamical equations for the transversal flow with the time coordinate playing the role of the axial coordinate in the steady flow (quasi-one dimensional-approximation), with the appropriate boundary conditions at the jet-ambient medium interface.

The code therefore solves the equations of relativistic magneto-hydrodynamics (RMHD) as a two-dimensional axisymmetric problem. The jet density $\rho_{\rm j}$, Lorentz factor $\gamma_{\rm j}$, and axial magnetic field component $B^{\rm z}_{\rm j}$ are initially constant across the jet. The toroidal magnetic field configuration is also fixed, and its profile is shown in \citet{Angles_2021}. As detailed in the paper, jets are injected in pressure equilibrium with the surrounding medium by matching the total pressure at the jet/ambient medium boundary, as described in \cite{Marti_2015} and \cite{Marti_2016}. The code incorporates the Synge relativistic gas equation of state \citep{Synge_1958} following approximations displayed in \cite{Choi_2010}.  The total jet power is defined\footnote{In the Eq.~\ref{eq: L_j} and throughout the whole paper, we absorb a factor $\sqrt{4 \pi}$ in the definition of the magnetic field.} as
\begin{equation}
    L_{\rm j} = \pi R_{\rm j}^2 \left(\rho_{\rm j} h_{\rm j} \gamma_{\rm j}^2 + \left(B^\phi_{\rm j}\right)^2\right)~v_{\rm j} \,,
    \label{eq: L_j}
\end{equation}
where $R_{\rm j}$ is the jet radius, $h_{\rm j}$ is the specific enthalpy and $v_{\rm j}$ is the jet velocity. The specific enthalpy is calculated as in \cite{Choi_2010},

\begin{align}
    h &= \dfrac{5c^2}{2\xi} + \left(2 - \kappa\right) c^2 \left[ \dfrac{9}{16}\dfrac{1}{\xi^2} + \dfrac{1}{\left(2 - \kappa + \kappa \mu\right)^2}\right]^{1/2}  \\
    & + \kappa c^2 \left[\dfrac{9}{16} \dfrac{1}{\xi^2} + \dfrac{\mu^2}{\left(2 - \kappa + \kappa \mu\right)^2}\right]^{1/2} \nonumber
\end{align}
with $\xi = \rho c^2 / p$ ($p$ being the gas pressure), $\kappa = n_{\rm p} / n_{\rm e}$ being the ratio of proton and electron number density and $\mu = m_{\rm p} / m_{\rm e}$. \\

For the particular configuration of the magnetic field used, the transversal equilibrium at injection leads to,
\begin{equation}
    p_{\rm j, 0} = p_{\rm a, 0} - \dfrac{\left(B^{\rm z}_{\rm j,0}\right)^2}{2}\,,
\end{equation}
where $p_{\rm j, 0}$, $p_{\rm a, 0}$ and $B^{\rm z}_{\rm j,0}$ are, respectively the mean jet pressure, the ambient pressure, and the jet axial magnetic field at injection. The fact that this pressure is independent of the toroidal magnetic field is a consequence of the particular profile of the toroidal magnetic field used, as it is discussed in \cite{Marti_2015}. As we will describe in the next section, all the parameters defining the jet and ambient medium for the base model used in this work correspond to model J4\_C of \cite{Angles_2021}.

The simulation box extends over a uniform grid described in cylindrical coordinates, $r$ and $z$, respectively, over $20~\rm{pc}$ and $2000~\rm{pc}$ with an initial jet radius of $R_{\rm j} = 1~\rm{pc}$. The number of computational cells in each direction of the grid is set to $1600 \times 10000$ and the axisymmetry of the problem imposes reflection at \textcolor{red}{$r = 0$}. As the mass-loading scenario shows promising results for FR\,I jets, the total power injected in the jet is set to $L_{\rm j} = 10^{43}~\rm{erg}\cdot\rm{s}^{-1}$. At the injection, the jet is purely leptonic with a fraction $X_{\rm e} = \rho_{\rm e}/\rho = 1.0$, where $\rho_{\rm e}$ and $\rho$ are, respectively, the leptonic and full rest-mass densities. As the jet propagates, the composition is set to evolve through star mass-loading. Table~\ref{tab:jet param} sums up the parameters used at the injection point (labeled with subscript $0$), either in the ambient medium or in the jet (respectively labeled $\rm a$ and $\rm j$). From those parameters, one can derive the mean gas pressure in the jet $p_{\rm j, 0}$, the axial field $B^{\rm z}_{\rm j, 0}$ and toroidal field $B^\phi_{\rm j, 0}\left(r\right)$, and the maximum toroidal field $B^\phi_{\rm m, j, 0}$. Such derivations can be seen in \cite{Angles_2021}. Full details of the parameters can be seen on their Table~1. We select the one labeled as J4\_C as our fiducial model, which corresponds to an average case of jet with a significant amount of kinetic energy flux ($F_{\rm k} \sim 28 \%$, see their Table~2).

\renewcommand{\arraystretch}{1.4}
\begin{table*}
    \caption{Jet, ambient medium, and stellar wind parameters used in the simulations, based on the J4\_C jet set up shown in \cite{Angles_2021}. From left to right, the different columns show the jet rest-mass density and flow Lorentz factor at injection, the ambient density and pressure at the jet base, the radius of the core of the galaxy atmosphere, the mass-load rate at the jet base, and the radius of the core of the stellar distribution.}
    \label{tab:jet param}
    \centering
    \begin{tabular}{l c c c c c c c c}
       \hline
       \hline
       Parameter & $\rho_{\rm j, 0}$ & $\gamma_{\rm j, 0}$& $\rho_{\rm a, 0}$ & $p_{\rm a, 0}$ & $r_{\rm c}$ & $Q_{\rm 0}$ & $r_{\rm c, s}$\\
       \hline
       Value  & $10^{-28}$ & $6$ & $10^{-24}$ & $10^{-7}$ & $500 - 1500$ & $10^{23}- 5 \times 10^{25}$  & $500 - 1500$\\
       Units & $\rm{g}\cdot\rm{cm}^{-3}$ & $-$ & $\rm{g}\cdot\rm{cm}^{-3}$ & $\rm{dyn}\cdot\rm{cm}^{-2}$  & $\rm{pc}$ & $\rm{g}\cdot\rm{yr}^{-1}\cdot\rm{pc}^{-3}$ & $\rm{pc}$\\
       \hline
    \end{tabular}
\end{table*}
\renewcommand{\arraystretch}{1}

\subsection{Ambient medium and mass-load}
\label{subsec:Ambient medium and stellar winds}

We consider an ambient medium pressure profile as described in \cite{Angles_2021} \citep[see also][]{Perucho_2014}, namely a power-law that reproduces a typical galactic atmosphere with
\begin{equation}
    p_{\rm a}\left(z\right) = p_{\rm a, 0} \left(1 + \left(\dfrac{z}{r_{\rm c}}\right)^2\right)^\alpha\,,
\end{equation}
where $r_{\rm c}$ represents the inner core radius, and $\alpha = -1.095$. Table~\ref{tab:jet param} shows the mass-load rate of ionized hydrogen (proton-electron) at the jet base, $Q_0$. Along the jet propagation axis, the average mass-load rate follows the profile
\begin{equation}
    Q\left(z\right) = Q_{\rm 0} \left(1 + \left(\dfrac{z}{r_{\rm c,s}}\right)^2\right)^{\alpha_{\rm s}}\,,
\end{equation}
with $r_{\rm c, s}$ being the stellar core radius and, as in \cite{Angles_2021}, we set $\alpha_{\rm s} = -1.095$. 
Here, we make the assumption that the mass-load profile follows the stellar distribution profile. This is motivated by the fact that mass-load should be driven by jet-star interactions, either from direct mass-load from stellar winds, and/or by mass-entrainment from the ambient medium due to the development of small-scale instabilities at the jet boundary triggered by stars entering and leaving the jet \citep{Perucho_2020}. In fact, from both analytic estimates and numerical simulations, it is known that stars can cross the jet without direct interaction, since the equilibrium point between the stellar wind and the jet flow is far beyond the star surface \citep{Komissarov_1994, Bosch_2012, Perucho_2017}. The total number of stars in a jet within its initial $10$ kiloparsecs is expected to be in the range of $10^{8}$ to $10^{9}$ \citep{Wykes_2014, Vieyro_2017}.  Finally, the stellar distribution profile used here is valid for giant elliptical galaxies, which are the typical hosts of FR I/II jets.

\begin{figure}
    \centering
    \includegraphics[width=\linewidth]{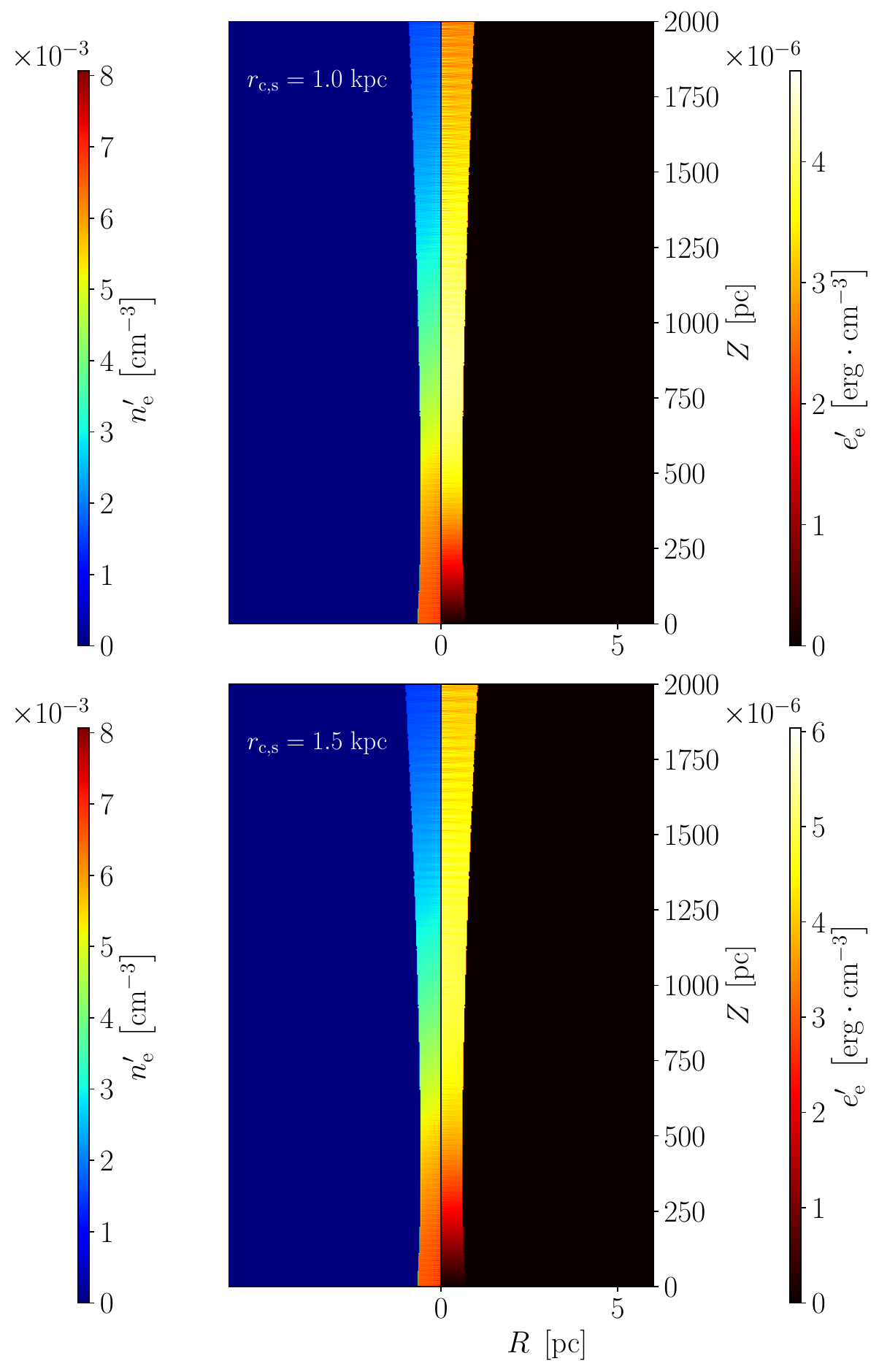}
    \caption{Jet simulations with $Q_0 = 5 \times 10^{24}~\rm{g}\cdot\rm{yr}^{-1}\cdot\rm{pc}^{-3}$}, the non-thermal electron density (left) and their associated non-thermal energy density (right), for the two stellar distributions $r_{\rm c, s} = 1.0~\rm{kpc}$ (up) and $r_{\rm c, s} = 1.5~\rm{kpc}$ (down). Axis are given in parsec units.
    \label{fig:dyn_exmpl}
\end{figure}

\subsection{Radiative transfer with the RIPTIDE code}
\label{subsec:Radiative transfer with the RIPTIDE code}

To compute synthetic synchrotron emission maps, we used the \texttt{RIPTIDE} code \citep{Fichet_2021, Fichet_2022}. We first consider a population of non-thermal electrons with the associated number density described by the power-law distribution 
\begin{equation}
    \mathrm{d}n^{\prime}_{\rm e}  = K \,\gamma^\prime_{\rm e}~^{-p} \, {\mathrm{d}\gamma^{\prime}_{\rm e}}\,, 
\end{equation}
where $K$ is the normalization constant of the power-law with spectral index $p = 2.2$, typical of mildly relativistic shock acceleration \citep{Ostrowski_2002, Lemoine_2003}, extending between $\gamma^\prime_{\rm e, min}$ and $\gamma^\prime_{\rm e, max}$.

The normalization constant and the minimum electron Lorentz factor can be derived as described in \cite{Gomez_1995}
\begin{equation}
    K = \left(\dfrac{e'_{\rm e} (p-2)}{m_{\rm e} c^2 (1 - C_{\rm E}^{2-p})}\right)^{p-1} \left( \dfrac{1 - C_{\rm E}^{1-p}}{n'_{\rm e} (p-1)}\right)^{p-2}\,,
\end{equation}

\begin{equation}
    \gamma^\prime_{\rm e, min} = \dfrac{1}{m_{\rm e} c^2} \dfrac{e^\prime_{\rm e}}{n^\prime_{\rm e}} \dfrac{p - 2}{p - 1} \dfrac{1 - C_{\rm E}^{1-p}}{1 - C_{\rm E}^{2-p}}\,,
\end{equation}
where $n^\prime_{\rm e}$ is the number density of the non-thermal electrons, and $e^\prime_{\rm e}$ its corresponding energy density. Finally, $C_{\rm E} = \gamma^\prime_{\rm e, max} / \gamma^\prime_{\rm e, min} = 10^3$, which is constant since we are neglecting radiative losses. This last assumption is supported by the fact that at one kiloparsec of the jet base, considering typical values as $\gamma^\prime_{\rm e, max} = 10^5$ and $B^\prime = 1~\rm{mG}$, the synchrotron radiative cooling timescale stays longer than the adiabatic one, with cooling lengths of several kiloparsecs, longer than the simulated jet. In the previous expression, both parameters $n^\prime_{\rm e}$ and $e^\prime_{\rm e}$ are retrieved from the numerical simulations, assuming that they are proportional respectively to the leptonic number density and the leptonic internal energy density \citep{Bottcher_2010, Mimica_2012, Fromm_2016}. In this work, the respective proportionality constants are set to $\xi_{\rm e} = 0.1$ and $\epsilon_{\rm e} = 0.1$  following previous procedures \citep{Gomez_1995, Mimica_2012, Fromm_2016, Fichet_2021}.

Accepting that our model is phenomenological, our approach is based on simple assumptions about the electron non-thermal population, leaving aside the details of the acceleration mechanisms responsible for the transfer of internal energy from the jet flow. Beyond this limitation, in the adopted model, $\gamma^\prime_{\rm e, min}$ is proportional to $e^\prime_{\rm e}/n^\prime_{\rm e}$, i.e., the (average) energy per (non-thermal) particle, which is reasonable. Additionally, this energy per (non-thermal) particle is proportional on its own to the internal energy per fluid particle. Furthermore, the fact that we fix the ratio between the maximum and minimum Lorentz factors of the power-law distribution, $C_E$, seems also plausible given the properties of the jets at the studied scales. Once this simple model was chosen, the welcome result is that, in many instances, $\gamma^\prime_{\rm e, min}$ (and the whole particle distribution) shifts to larger values downstream the jet (due to the dissipation of kinetic energy), which is on the basis of the positive radio-optical offsets obtained in the present work.

The synchrotron emissivity $j_\nu^\prime $ and absorption $\alpha_\nu^\prime$ coefficients are computed in the jet frame according to the approximations shown in \cite{Katarzynski_2001}. Transformation into the observer frame is accounted for through the following relations \citep{Rybicki_1979}

\begin{align}
    j_{\nu} &= \delta^2 j^\prime_{\nu} \,, \\
    \alpha_{\nu} &= \delta^{-1} \alpha^\prime_{\nu} \,.
\end{align}
These relations, which are valid for a steady jet flow, depend on the Doppler factor $\delta = \left(\gamma_{\rm j} \left(1 - \beta_{\rm j} \cdot \cos\left(\theta_{\rm obs}\right)\right)\right)^{-1}$, with $\beta_{\rm j} = v_{\rm j} / c$ and $\theta_{\rm obs}$ the angle between the direction of the jet axis and the line of sight. Now, the resulting specific intensity at a given cell with index $i$ along a line of sight for an observer in the absolute frame is
\begin{equation}
    I_{\nu, \,\rm i} = I_{\nu, \, i-1} \exp{\left(-\tau_{\nu, \, i}\right)} + S_{\nu, \, i} \left(1 - \exp{\left(-\tau_{\nu, \, i}\right)} \right),
    \label{eq:int2}
\end{equation}
with $I_{\nu, \, i-1}$ being the incident intensity on the cell $i$, $\tau_{\nu, \, i}$ is the optical depth due to synchrotron self-absorption and $S_{\nu, \, i} = j_{\nu, \, i} / \alpha_{\nu, \, i}$ is the synchrotron source function, considered constant inside the cell. For each line of sight, the emergent specific intensity is then $I_\nu = I_{\nu, \, N}$, with $N$ being the total number of jet cells crossed by the line of sight. The incident specific intensity entering the jet is set to $I_{\nu, \, 0} = 0$.

The computed synchrotron intensity $I_\nu$ is converted into a synchrotron flux $F_\nu$
\begin{equation}
    F_\nu = \left(\dfrac{S_{\rm e}}{D_{\rm L}^2}\right) \cdot \left(1 + z\right) \cdot I_\nu\,,
\end{equation}
with $S_{\rm e}$ being the emitting surface and $D_{\rm L}$ the luminosity distance that depends on the shift of the source $z$. We neglect the impact of the light travel time delays, as we are interested in the steady state of the source \citep[e.g.,][]{Fichet_2022}.

We apply an optical flux selection criterion based on the sensitivity of the \emph{Gaia} mission to the integrated optical flux, as computed from our \texttt{RIPTIDE} model. Specifically, in the AB magnitude system, the \emph{Gaia} telescope has a flux sensitivity of $F_{\rm AB, Gaia} = 10^{-4}~\rm{Jy}$ at a $3\sigma$ confidence level, according to the DR3 catalog \citep{Gaia_2021}. This means that only sources with an optical flux above this threshold will be considered for analysis. Similarly, we apply a radio flux threshold at the milli-Jansky (mJy) level. However, this radio flux cut does not significantly affect our results, as the optical flux selection is more restrictive and is thus the primary factor that determines which sources are included in our study. For our continuous radio and optical emission maps, we define the coordinates in terms of the centroid position (maximum flux position) of the emission, and we build the angle between the radio to optical peak, denoted as $\Psi$ (as in \citetalias{Plavin_2019}).


\section{Radio-optical positive offsets}
\label{sec:Results}

\subsection{Energy dissipation}
\label{subsec:Energy dissipation}

\begin{figure}
    \centering
    \includegraphics[width=\linewidth]{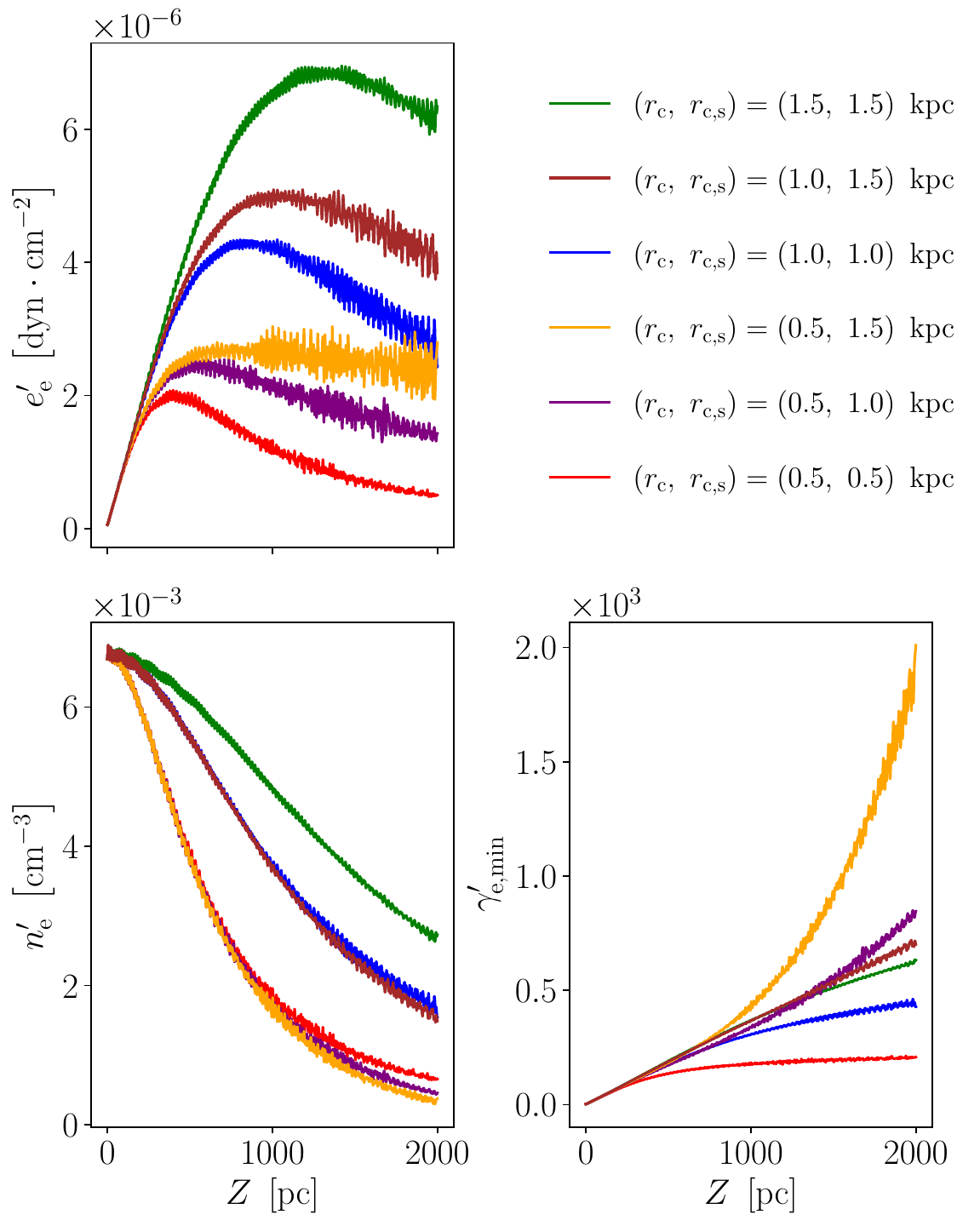}
    \caption{Axial distance dependence of the non-thermal electron energy density $e^\prime_{\rm e}$, density $n^\prime_{\rm e}$ and minimum Lorentz factor $\gamma^\prime_{\rm e,min}$ evaluated along the jet inner axis, given in parsecs. In all cases, an average mass-load rate of $Q_0 = 5 \times 10^{23}~\rm{g}\cdot\rm{yr}^{-1}\cdot\rm{pc}^{-3}$ was used}.
    \label{fig:evolv_rcs}
\end{figure}

At the scale relevant to this study, jets are accelerated and expect that magnetic acceleration has already taken place \citep[e.g.][]{Ricci_2024}. Thus, the jet must carry a non-negligible amount of kinetic and internal energies. Here, we chose the simulation labeled J4\_C from \cite{Angles_2021} as it represents a mildly kinetic jet, positioned between the two cases studied in their paper: a kinetically dominated jet (J4\_A) and an internal energy dominated jet (J8\_X). Our choice of J4\_C allows for the exploration of an intermediate scenario. Additionally, we tested the energy dissipation behavior across these three cases and verified that while the J8\_X configuration shows no energy dissipation, both J4\_C and J4\_A exhibit similar levels of energy dissipation due to their kinetic properties, which is consistent with \cite{Angles_2021}. This also aligns with the conclusions of \cite{Bowman_1996} and the findings in \cite{Angles_2021}, which highlight that energy dissipation occurs primarily in kinetically (dominated) jets due to mass-loading processes, increasing internal energy. \\

 To study the presence and characteristics of radio-optical offsets, we first investigate the impact of different stellar distributions by setting $r_{\rm c,s} = 1.0~\rm{kpc}$ and $1.5~\rm{kpc}$. Figure~\ref{fig:dyn_exmpl} shows 2D maps of the non-thermal electron number density $n^\prime_{\rm e}$ (left) and the non-thermal electron energy density $e^\prime_{\rm e}$ (right), for an average mass-load rate of $Q_0 = 5 \times 10^{23}~\rm{g}\cdot\rm{yr}^{-1}\rm{pc}^{-3}$. For such a low $Q_0$ value, the jet appears still collimated ($\theta_{\rm open} \sim 0^\circ$) while a local increase of the non-thermal energy density can be seen due to mass-loading (with a maximum value at axial distance $Z \sim 0.9~\rm{kpc}$ and $Z \sim 1.2~\rm{kpc}$, respectively), which is related to the value of $r_{\rm c,s}$. Due to the incorporation of protons, the ratio of the number of protons per electron $\kappa$ increases with $Z$.

The ambient medium can also play a role in the radiative jet profile, allowing jet expansion within steep pressure profiles or the opposite. To investigate the dependence of $n^\prime_{\rm e}$ and $e^\prime_{\rm e}$ with the ambient gas density and stellar distributions, we compute the electron number density $n^\prime_{\rm e}$ and the associated non-thermal energy density $e^\prime_{\rm e}$ profiles for various values of $r_{\rm c}$ and $r_{\rm c,s}$. Figure~\ref{fig:evolv_rcs} shows various profiles of those variables for different sets of $r_{\rm c}$ and $r_{\rm c,s}$, for a fixed value of $Q_0 = 5 \times 10^{23}~\rm{g}\cdot\rm{yr}^{-1}\cdot\rm{pc}^{-3}$. As shown, the position of maximum energy dissipation (maximum of the non-thermal energy density) evolves according to the value of the core radii. In parallel, the electron number density decreases at different rates. We suggest that a positive radio-optical offset is likely to occur when the position of the maximum energy dissipation is away from the jet base, due to the evolution of $\gamma_{\rm e,min}$ along the jet axis. Indeed, this displacement can be attributed to the fact that, as energy dissipation peaks further away from the jet base, the resulting emission at different wavelengths becomes spatially separated: both optical and radio emission would be localized along the jet axis, although at different axial positions.

\begin{figure}
    \centering
    \includegraphics[width=\linewidth]{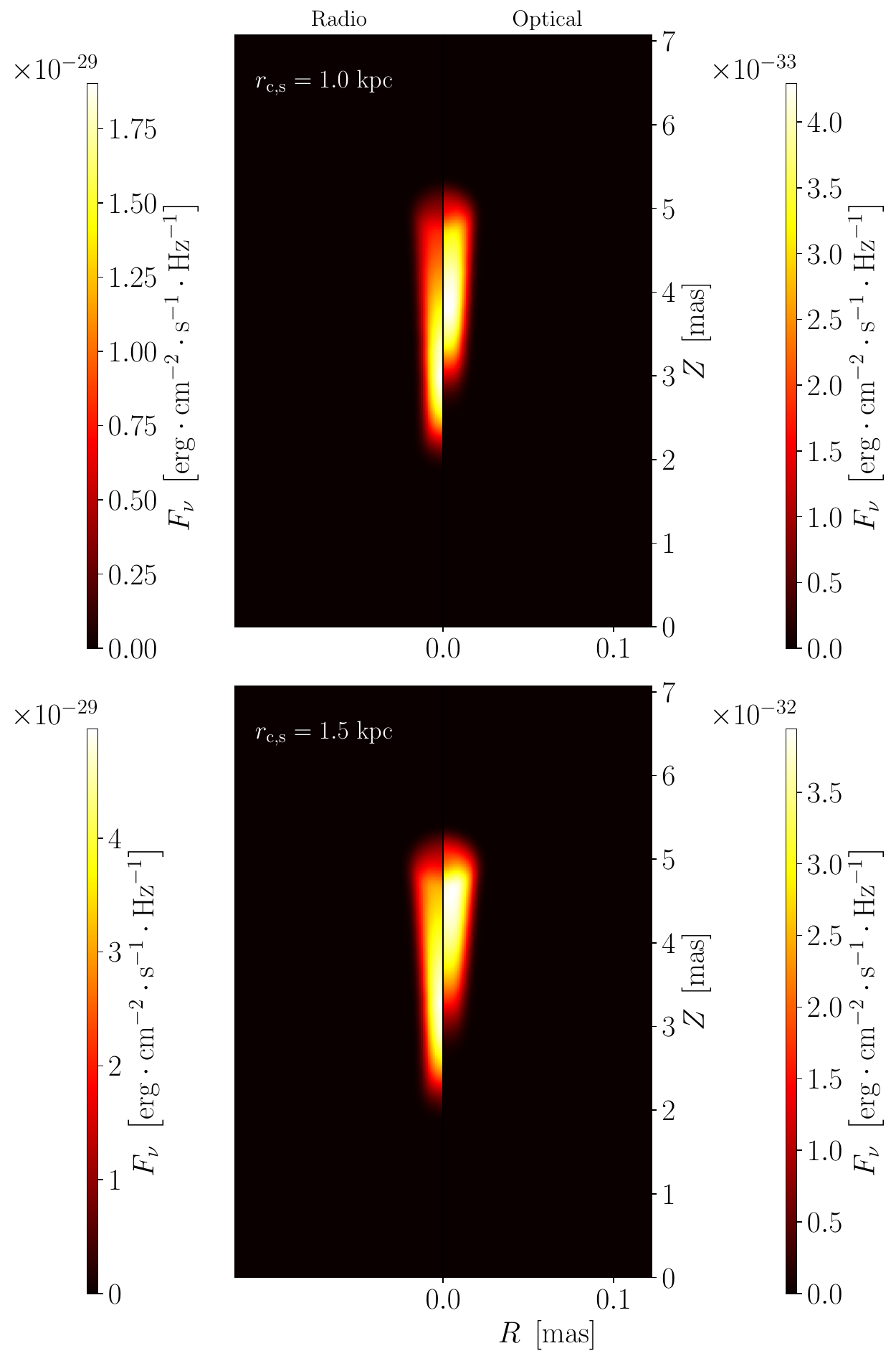}
    \caption{Synthetic synchrotron flux maps in the radio ($\nu = 3 \times 10^{11}~\rm{Hz}$, left) and in the optical band ($\nu = 5 \times 10^{14}~\rm{Hz}$, right) from simulations shown on Figure~\ref{fig:dyn_exmpl}. We respectively display the results for the two stellar distributions $r_{\rm c,s} = 1.0~\rm{kpc}$ (up) and $r_{\rm c, s} = 1.5~\rm{kpc}$ (down) for a viewing angle $\theta_{\rm obs} = 10^\circ$, redshift $z = 1.4$, a fixed $r_{\rm c} = 1.0~\rm{kpc}$ and a fixed average mass-load of $Q_0 = 5.0 \times 10^{23}~\rm{g}\cdot\rm{yr}^{-1}\cdot{pc}^{-3}$}.
    \label{fig:em_exmpl}
\end{figure}

To study the presence and characteristics of radio-optical offsets, we produce synthetic images, computed using \texttt{RIPTIDE} on the jet simulations. If not stated otherwise, we construct emission maps for two frequencies in the observer's frame $\nu_{\rm obs} = 4.3 \times 10^{10}$~/~$5 \times 10^{14}~\rm{Hz}$ (corresponding to typical VLBI/optical frequency) and for a given pair of jet viewing angle $\theta_{\rm obs}$ and shift $z$, and thus evaluated in the jet frame. The axis of the emission maps is given in milliarcseconds to be compared to observations \citepalias{Plavin_2019}. For reference, Figure~\ref{fig:em_exmpl} shows emission maps associated with the simulations shown on Figure~\ref{fig:dyn_exmpl}, with a radio emission map shown on the left and the optical one on the right. Optical emission maps display a dimmer jet at the base, which becomes brighter with distance due to the increase of $\gamma^\prime_{\rm e, min}$ triggered by mass-loading. A clear radio-optical positive offset $d_{\rm app}$ is seen in both cases, with values ($\sim 0.4$ and $\sim 1~\rm{mas}$, respectively) depending on the choice of the stellar distribution ($r_{\rm c,s}$). 

To study the radio-optical positional offsets, $d_{\rm app}$, for a range of $\dot{M}$ values, we measure the offset values for a set of emission maps by fixing the jet observation angle, $\theta_{\rm obs} = 5^\circ$, and redshift, $z = 1$. These values are the medians derived by the authors from the collection of quasars in \citetalias{Plavin_2019}. 

\begin{figure}
    \centering
    \includegraphics[width=\linewidth]{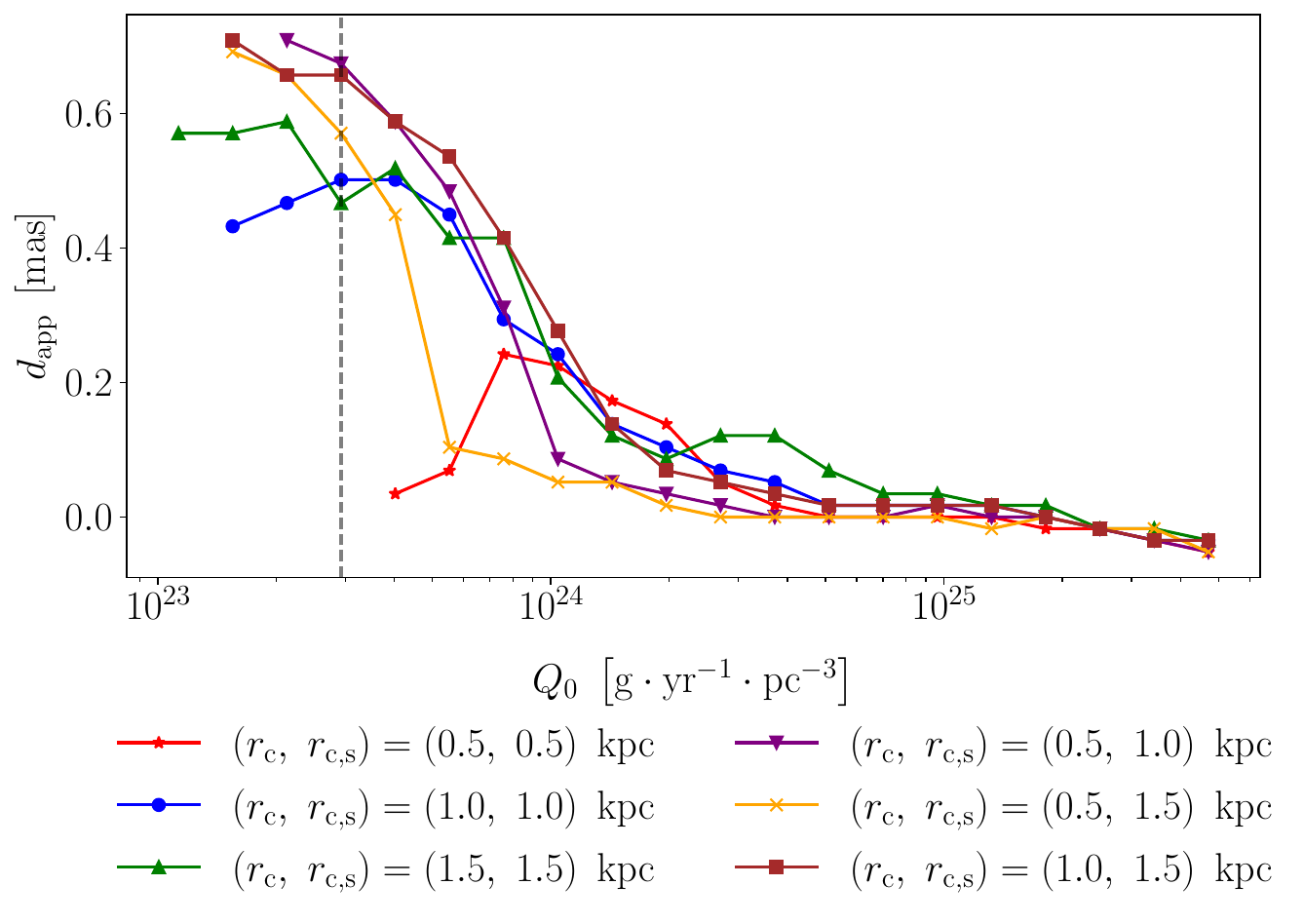}
    \caption{Dependence of the apparent radio-optical offset $d_{\rm app}$ with the average mass-load $Q_0$} for various ambient medium and stellar core radii ($r_{\rm c}$ and $r_{\rm c,s}$). Offsets on the right side of the vertical black dashed lines represent minimum offsets detectable by \emph{Gaia}, according to our selection criteria.
    \label{fig:range_q0}
\end{figure}

Figure~\ref{fig:range_q0} shows the variation of $d_{\rm app}$ with $\dot{M}$ for the same set of core radii $r_{\rm c}$ and $r_{\rm c,s}$. The radio-optical offset decreases with increasing values of mass-loading rates. For low $\dot{M}$, the energy dissipation not only occurs at a lower rate but also favors the optical emission peak to appear at a larger distance from the jet base. This configuration thus results in higher values of the radio-optical offset, as dissipation takes place in an extended region, as opposed to strong mass-loading, in which deceleration is fast and therefore dissipation takes place at the jet base. Therefore, the tendency for $d_{\rm app}$ separation to decrease with $\dot{M}$ is mainly due to the increase of the mass-load in the jet per unit of distance. 

Consequently, the location of maximum energy dissipation begins to drift closer toward the jet base, causing the radio and optical centroid emissions to converge closer to the origin. At high $\dot{M}$ values, this convergence intensifies, and the radio-optical emission offset approaches zero, indicating a minimal spatial separation between these emissions close to the jet base. As a competing effect, the optical flux increases with increasing values of $\dot{M}$ making the jet detectable by \emph{Gaia}. The vertical dashed line represents, in that case, the lowest $\dot{M}$ value that allows detection by \emph{Gaia}. Therefore, one needs a proper range of $\dot{M}$ that allows a high enough optical flux for detection and also extended dissipation. Overall, our results suggest the existence of an ideal average mass-load range, allowing us to observe a detectable and non-zero radio-optical offset spanning from $5 \times 10^{23}$ to $5 \times 10^{24}~\rm{g}\cdot\rm{yr}^{-1}\cdot\rm{pc}^{-3}$.

Regarding the role of the host galaxy's gas and stellar distributions, we observe that increasing $r_{\rm c,s}$ generally results in larger values of $d_{\rm app}$ indicating that the peak energy dissipation occurs further from the jet base. Conversely, smaller values of $r_{\rm c}$ favor fast jet expansion, causing the capture of more stars inside the jet and favoring jet deceleration, which in turn causes the peak energy dissipation to move closer to the jet base. It appears that the bigger is the galaxy, the larger the radio-optical offset value can reach, putting aside observational biases on the viewing angle and redshift. 

\subsection{Distribution of radio-optical offsets}
\label{subsec:Radio-optical shift}

Interestingly, from all our simulated cases, we find a typical offset value close to the median one referred in \citetalias{Plavin_2019} with $d_{\rm app} \sim 0.7~\rm{mas}$ typical quasar parameters with $\theta_{\rm obs} = 5^\circ$ and $z = 1$ (see Figure~\ref{fig:range_q0}). However, the reliability of this result should be taken with caution since our analysis considers all the jets that are detectable, without accounting for the angular resolution limits of \emph{Gaia}. To extend this study to a larger range of observation angles and redshift, and simulate observations of jets, we randomize a set of $1000$ sources with a uniform distribution of observation angles from $1^\circ$ to $30^\circ$, and a uniform distribution of different redshifts from $0.5$ and $2$. The choice of uniform distributions is solely based on avoiding introducing additional biases.

\begin{figure}
    \centering
    \includegraphics[width=\linewidth]{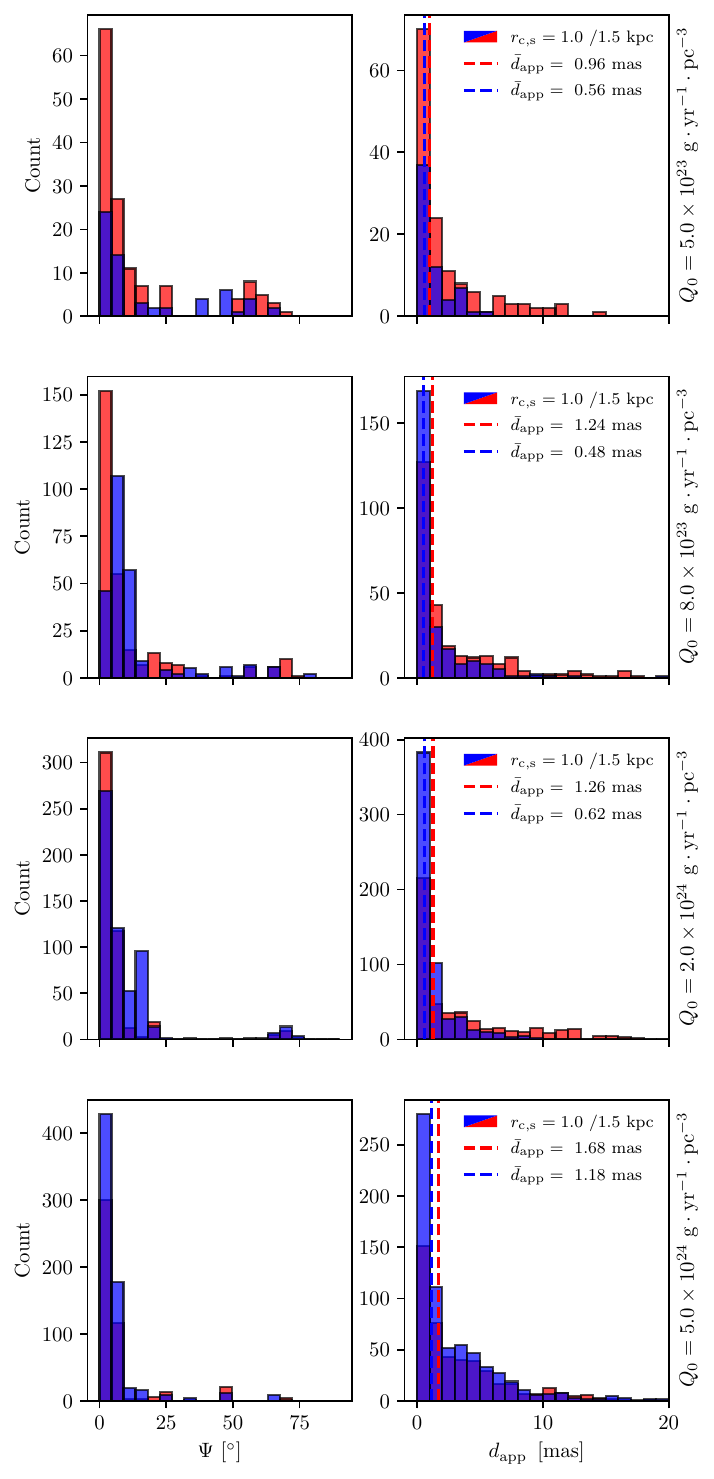}
    \caption{Histograms of the radio-optical angle ($\Psi$, left) and apparent offset ($d_{\rm app}$, right) for the set of randomized synthetic emission maps. Each row represents histograms for a given average mass-load $Q_0$} and for a given star distribution ($r_{\rm c, s} = 1.0~\rm{kpc}$ in blue and $r_{\rm c,s} = 1.5~\rm{kpc}$  in red; dark blue represents the overlap). Vertical dashed lines in the right column panels represent the associated median radio-optical offset.
    \label{fig:histo_offset}
\end{figure}

For each set of sources, we use $8$ values of $Q_0$ ranging from $5 \times 10^{23}$ to $5 \times 10^{24}~\rm{g}\cdot\rm{yr}^{-1}\cdot\rm{pc}^{-3}$. Additionally, we test the two cases of different stellar distributions $r_{\rm c, s} = \left(1.0,~1.5\right)~\rm{kpc}$ for a common ambient medium ($r_{\rm c} = 1.0~\rm{kpc}$), which allow us to study solely the impact of the stellar distribution. This choice is based on the previous results shown in Figure~\ref{fig:range_q0}. The results are shown in Figure~\ref{fig:histo_offset}, where the left column represents the distributions of the angles between the optical and radio peak directions $\Psi$\footnotemark  as measured from the jet base (see Sec.~\ref{sec:Introduction}) and the right column shows the distributions of offsets $d_{\rm app}$, for different average mass-load rates and stellar core sizes. We limit the results shown to the cases for which the optical emission can be detected by \emph{Gaia} according to its sensitivity. Even if we consider a straight (linear) jet here, one could simulate bent jets to study how the peak positions of the optical emission evolved according to Doppler boosting, which may also cause a small, non-zero $\Psi$. Taking into account the limited angular resolution of \emph{Gaia} will reduce the presence of non-zero $\Psi$ values.

In all the cases, positive offsets have been obtained with $d_{\rm app} \gtrsim 0~\rm{mas}$ and $\Psi \sim 0^\circ$. A stellar distribution with $r_{\rm c,s} = 1.0~\rm{kpc}$ leads to smaller median offsets, $d_{\rm app}$, (as seen on in histograms of Figure~\ref{fig:histo_offset}), as already discussed. 

\begin{figure*}
    \centering
    \includegraphics[width=0.95\linewidth]{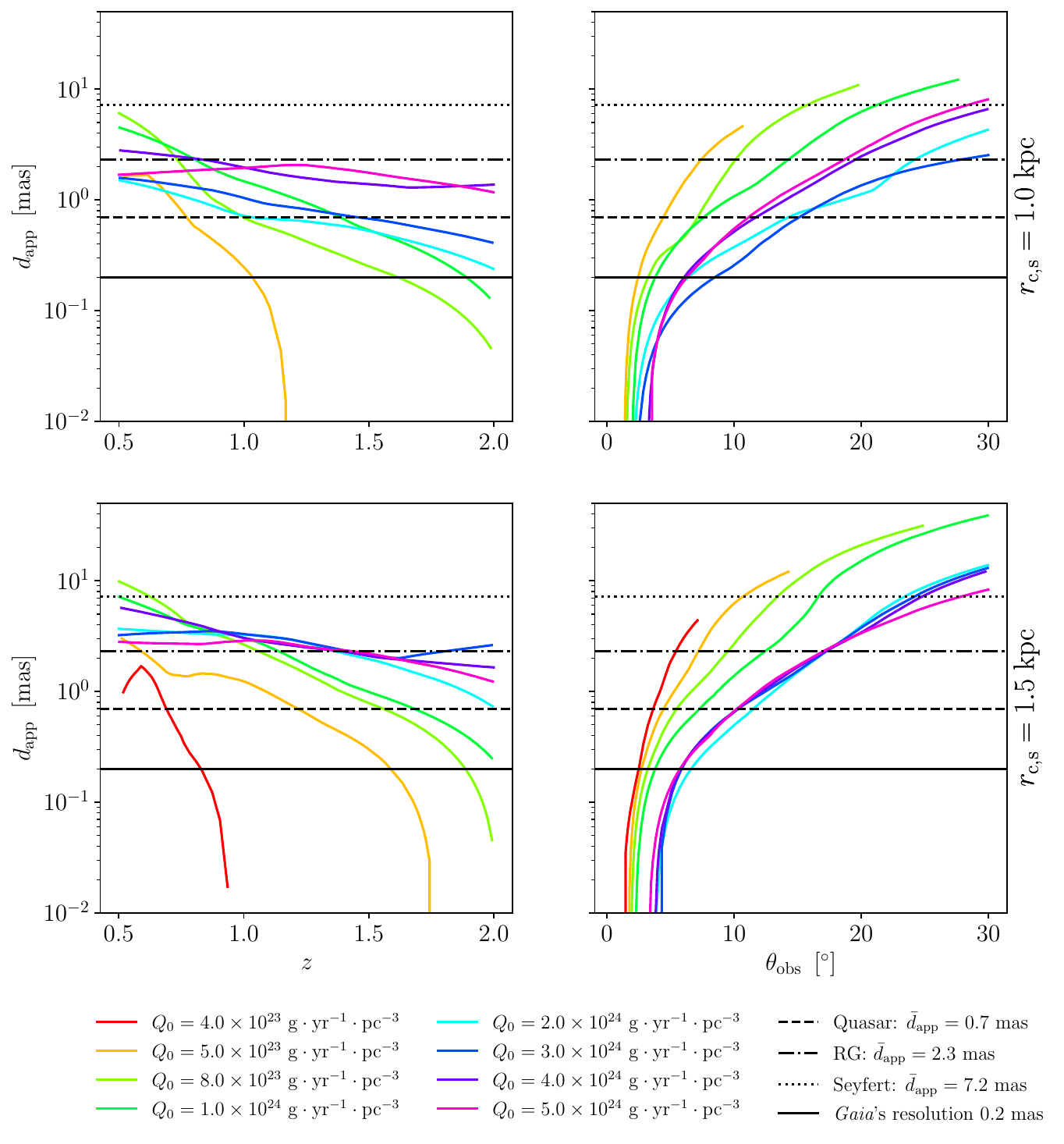}
    \caption{Dependence of the apparent radio-optical offset $d_{\rm app}$ with the redshift $z$ (left column) and with jet observation angle $\theta_{\rm obs}$ (right column). For various average mass-load rates $Q_0$} (see legend), solid lines represent the average across our randomized set of jets. Results are displayed for the two stellar distributions tested ($r_{\rm c,s} = 1.0~\rm{kpc}$ first row and $r_{\rm c,s} = 1.5~\rm{kpc}$ second row), with a fixed gas distribution of $r_{\rm c} = 1.0~\rm{kpc}$. The horizontal black lines illustrate medians $d_{\rm app}$ referred in \citetalias{Plavin_2019} for quasars ($0.7~\rm{mas}$, dashed line), radio galaxies ($2.3~\rm{mas}$, dash dotted line) and Seyfert $2$ ($7.2~\rm{mas}$, dotted line) in case of significant offset detection. We also represent here the typical \emph{Gaia}'s angular resolution as a solid black line for qualitative comparisons.
    \label{fig:d_app_evolv_z}
\end{figure*}

Figure~\ref{fig:d_app_evolv_z} presents the dependency of the offset values $d_{\rm app}$ for jet cases detectable by \emph{Gaia}, extracted from our dataset of $1000$ synthetic jet emission maps, with redshift $z$ (left column) and viewing angle $\theta_{\rm obs}$ (right column). Each plot includes solid lines that illustrate the average radio-optical offset for different mass-loading profiles in each redshift ($0.1$) or viewing angle ($1.0$°) bins, as detailed in the legend, joined by the actual \emph{Gaia}'s angular resolution. The top and bottom rows of each column correspond to star distributions with core radii $r_{\rm c,s}$ of $1.0~\rm{kpc}$ and $1.5~\rm{kpc}$, respectively. This layout highlights the effects of observational biases, such as redshift and observation angle, on the apparent displacement for a constant mass-loading value. 

\footnotetext{Note that our models are cylindrically symmetric, non-zero radio-optical angles $\Psi$ are unreliable and are a measure of the limitations of our procedure to detect the optical emission maxima.}

Indeed, due to the selection effect imposed by \emph{Gaia}'s sensitivity, a larger radio-optical offset will be detectable for low redshift sources and large observation angles. Physical impact of the mass-loading profile can be seen with great differences between $5 \times 10^{23}$ and $5 \times 10^{24}~\rm{g}\cdot\rm{yr}^{-1}\cdot\rm{pc}^{-3}$, where the offset $d_{\rm app}$ can vary by a factor $10$ (for a constant redshift, or a constant observation angle). For large mass-loading values, the average profiles of $d_{\rm app}$ seem to converge as the effect of mass-loading saturates. Assuming the mass-load is mainly driven by stellar winds, we can relate the flux detectability of jets by \emph{Gaia} to the distribution of stars in the host galaxy. Jets that experience lower mass-loading are more likely to be detected for larger stellar core radii as more stars interact with the jet. For example, a jet encountering an average mass-load of approximately $Q_0 = 5 \times 10^{24}~\rm{g}\cdot\rm{yr}^{-1}\cdot\rm{pc}^{-3}$ is detectable when the core radius of the star distribution, $r_{\rm c,s}$, is about $1.5~\rm{kpc}$, but not at $1.0~\rm{kpc}$. In general, increasing the core radius, $r_{\rm c,s}$, leads to a larger apparent distance, $d_{\rm app}$, and a greater optical flux for a given average mass-load rate, $Q_0$.

 On Figure~\ref{fig:d_app_evolv_z}, we also highlight median offsets for quasars, $d_{\rm app} = 0.7~\rm{mas}$, radio galaxies, $2.3~\rm{mas}$, and Seyfert $2$ galaxies, $7.2~\rm{mas}$, which is in remarkable agreement with the observational values given by \citetalias{Plavin_2019} from their sample. These median values are particularly interesting, as they underline the dependence of the radio-optical offset with the Doppler effect. Indeed, from our results, quasar objects can be observed at larger redshifts (between $1$ and $1.5$) and low observation angles $0^\circ$ to $10^\circ$, while radio galaxies should be observed at lower redshifts and larger observation angles, which is even more pronounced for Seyfert $2$ galaxies. Those findings align perfectly with both \citepalias{Plavin_2019} and our current knowledge on AGN classification \citep[e.g.][]{Netzer_2015}. 

Regarding the effects of the stellar distribution, we note that a decrease in the stellar core radius from $r_{\rm c,s} = 1.5~\rm{kpc}$ to $r_{\rm c,s} = 1.0~\rm{kpc}$ shifts the apparent radio-optical offset to smaller values (requiring an increase by several degrees in the viewing angle to recover the original apparent offset). A larger stellar core holds the dissipation of energy over longer distances, thus making the peak energy dissipation to move away from the jet base. Assuming a range of average mass-load profiles, the different jet profiles can exhibit varied deceleration characteristics, which influence the optimal observation angles for detecting similar displacement values.
It is important to note that since we do not impose any dependence on the average mass-load rate with galaxy redshift, the differences in detectability and $d_{\rm app}$ values are primarily due to selection biases related to the observational setup. 

\subsection{Color magnitude of the jet}
\label{subsec:Color magnitude of the jet}

\begin{figure}
    \centering
    \includegraphics[width=0.78\linewidth]{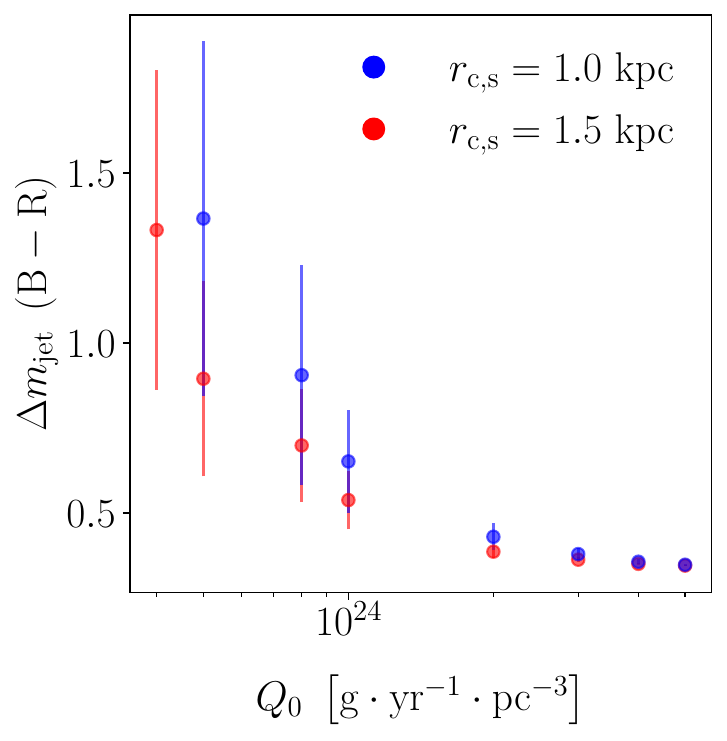}
    \caption{Dependence of the difference average magnitude $\Delta m_{\rm jet}$ between blue (B) and red (R) bands with the mass-load rate $Q_0$}. Results are displayed for the two stellar distributions tested, $r_{\rm c,s} = 1.0~\rm{kpc}$ in blue, and $1.5~\rm{kpc}$ in red with associated $1\sigma$ standard deviation uncertainty bars as we average over the $\theta_{\rm obs}$ and $z$ distribution.
    \label{fig:dm_jet}
\end{figure}

\begin{figure}
    \centering
    \includegraphics[width=0.8\linewidth]{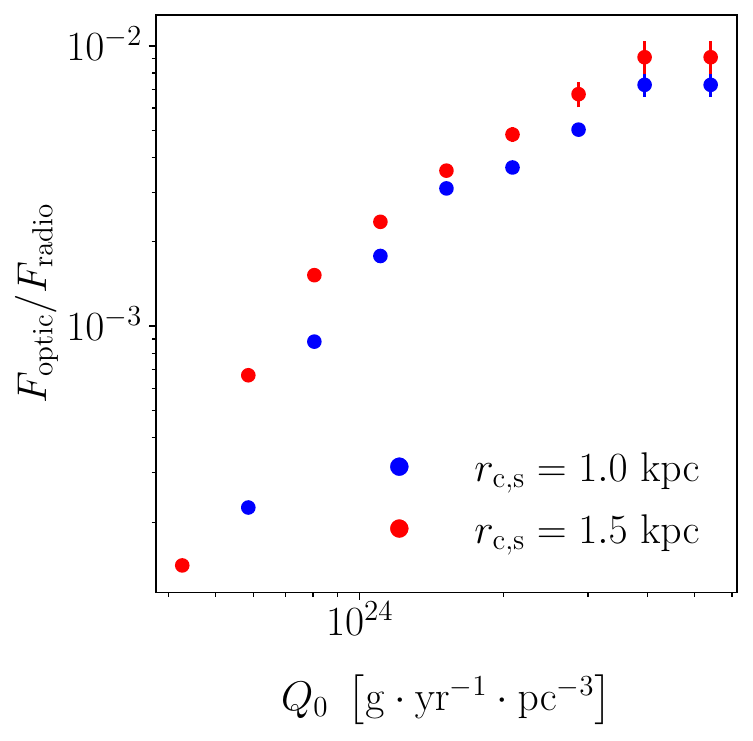}
    \caption{Dependence of the average optical over radio fluxes $F_{\rm optic} / F_{\rm radio}$ (as measured respectively at $\nu = 3 \times 10^{10}~\rm{Hz}$ and $\nu = 5 \times 10^{14}~\rm{Hz}$), with the average mass-load rate $Q_0$}. Results are displayed for the two stellar distributions tested, $r_{\rm c,s} = 1.0~\rm{kpc}$ in blue, and $1.5~\rm{kpc}$ in red with associated $1\sigma$ standard deviation uncertainty bars as we average over the $\theta_{\rm obs}$ and $z$ distribution.
    \label{fig:ratio_opt_rad}
\end{figure}

\citetalias{Plavin_2019} show how the presence of a bluer object is generally linked to the presence of an accretion disk (and of a so-called big blue bump component), which translates into a negative radio-optical offset. Indeed, in those cases, the optical centroid is shifted towards the base of the jet. Therefore, redder objects seem to be linked to objects showing positive radio-optical offsets and a relatively less powerful accretion disk component.
Although the inclusion of an accretion disk component is beyond the scope of this paper, one can estimate the difference magnitude $\Delta m_{\rm jet}$ between the blue and red bands of the \emph{Gaia} telescope in the jet. This parameter is estimated as follows,
\begin{equation}
    \Delta m_{\rm jet} = -2.5 \log_{\rm 10} \left(\dfrac{F_{\rm B}}{F_{\rm R}}\right) \,,
\end{equation}
where $F_{\rm R}$ and $F_{\rm B}$ are respectively the fluxes as measured in the red and blue bands of \emph{Gaia}. For simplicity, we consider the wavelength centered in the middle of each band, respectively, at $\rm{R} \sim 850~\rm{nm}$ and $\rm{B} \sim 550~\rm{nm}$ \citep{Jordi_2010}. Consequently, a larger positive value of $\Delta m_{\rm jet}$ indicates a redder jet.

Figure~\ref{fig:dm_jet} shows the variation of $\Delta m_{\rm jet}$ as a function of the average mass-load rate, for the two-star core radii $r_{\rm c,s} = 1.0~\rm{kpc}$ and $1.5~\rm{kpc}$. The results for the two stellar core radii are almost indistinguishable, with red fluxes dominating over blue fluxes. In both cases, the red excess tends towards a minimum increasing $Q_0$ since larger mass-load rates increase the amount of dissipated energy and, hence, the optical flux. 

Knowing that smaller $Q_0$ values lead to larger radio-optical offsets, this result is under the correlation with redder objects, as observed and discussed in \citetalias{Plavin_2019}. Putting aside the accretion disk contribution in the overall magnitude, our results show that redder objects should imply lower mass-loading rates, and observable (discernible) positive radio-optical offsets according to our results. For larger values of $Q_0$, the difference in color magnitude saturates as the increase of $Q_0$ leads $d_{\rm app}$ converging to zero. 

\cite{Petrov_2017} underline that the optical centroid detected by \emph{Gaia} is dominated by compact, jet-related emission rather than by diffuse stellar light from the host galaxy. However, if a significant part of the mass-loading in the jet originates from stellar winds, the presence of a stellar population could impact the radio-optical offset and potentially alter the magnitude estimates in our analysis. Since the jet’s optical emission is highly compact compared to the extended stellar distribution, the impact of the stellar luminosity on the optical centroid position is expected to be limited. A detailed inclusion of this component falls outside the scope of this paper, but future work could aim to consider specific stellar types and distributions, as the influence on \emph{Gaia}’s optical centroid may vary in galaxies with substantial mass-loading from stellar sources. Still, our results are consistent with the observational results from \citetalias{Plavin_2019}, even though we cannot compare our difference magnitude with the \emph{Gaia} color observed, due to both physical (accretion disk and stellar population not simulated here) and instrumental aspects.


\section{Discussion}
\label{sec:Discussion}

\subsection{The mass-loading scenario}
\label{subsec:the mass-loading scenario}

We use RMHD numerical simulations of jets including source terms that account for mass-load, following previous work by \cite{Angles_2021}. With these simulations, we have studied the possible role of dissipation in the presence of radio-optical offsets by means of the radiative transfer code \texttt{RIPTIDE}. We have chosen a fixed jet power of $L_{\rm j} = 10^{43}~\rm{erg}\cdot \rm{s}^{-1}$, a typical value for FR\,I jets \citep{Perucho_2020}, for which the proposed scenario shows promising results. The basic idea lies in the fact that efficient dissipation is known to largely coincide with the region in which jet deceleration takes place (see Figure~\ref{fig:dyn_exmpl}), as shown by \cite{Laing_2014}. Due to the conversion of kinetic energy into internal one, the minimal Lorentz factor of the non-thermal electrons increases with distance according to their non-thermal energy density (see Figure~\ref{fig:evolv_rcs}). The presence of non-thermal emission from a population of accelerated particles is suggested from optical polarimetry observations \citep{Kovalev_2020}, and we do derive positive radio-optical offsets from our synchrotron maps. Jet-star interactions are unavoidable in the jets \citep{Wykes_2014, Vieyro_2017}, and they may play a relevant role in jet deceleration according to several theoretical \citep[e.g.][]{Komissarov_1994, Perucho_2020} and observational works \citep{Mingo_2019}. Although the setup differs, our radio emission maps bear similarities to those observed in \cite{Mimica_2009}, particularly in the case of stationary, pressure-matched jets relative to the ambient medium. Despite not accounting for radiative losses due to their specific parameters, the authors demonstrate that such losses shorten the emission morphology, while the centroid remains in the same position. Nevertheless, at the scale studied here, we consider that the leading cooling process is driven by adiabatic expansion. It should be noted that jet bends or other dissipation processes such as shocks or turbulence could imply an additional shift of the radio-optical centroid beyond those analyzed here based on the jet mass-load. \\

Since our model to explain radio-optical offsets is based on the dissipation of kinetic energy, we have chosen one of the models analyzed in \cite{Angles_2021} with a large fraction of kinetic flux as the base jet model of our study. At the relevant scales for this study, jets appear to be relativistic and tend to carry a significant amount of kinetic energy, consistent with theoretical and numerical findings. These findings indicate that beyond the acceleration region, typically a few parsecs from the jet base—jets are predominantly kinetically dominated \citep{Vlahakis_2004, Komissarov_2009, Komissarov_2012, Ricci_2024}. This aligns with the dissipation mechanisms required to produce the observed radio-optical offsets. Here we show that, for a certain regime of average mass-load rates, ranging from $5 \times 10^{23}$ to $5 \times 10^{24}~\rm{g}\cdot\rm{yr}^{-1}\cdot\rm{pc}^{-3}$, the dissipation produced in the jet may result in radio-to-optical peak offset values between $d_{\rm app} \simeq 0.4~\rm{mas}$ and $10~\rm{mas}$, which is consistent with observations of different types of AGN (see Figure~\ref{fig:range_q0} for application to quasar type). Indeed, jets at large redshift and small observation angles are consistent with the quasar population. In contrast, smaller redshift and larger observation angles are consistent with radio galaxies and Seyfert types \citepalias{Plavin_2019}. In this sense, the impact of Doppler boosting appears crucial to understand the characteristics of radio-optical offsets \citep{Secrest_2022}.

Our results indicate the presence of a higher average mass-load rate for radio galaxies and Seyfert~$2$ galaxies. The latter type is known to host star-forming regions, that could enhance locally the average stellar mass-loss rate \citep{Rodriguez_1987}, potentially affecting the radio-optical offset. Our color magnitude estimation indicates that detectable offsets are correlated with relatively red objects (see Figure~\ref{fig:dm_jet}), which is also consistent with observations \citep[\citetalias{Plavin_2019};][]{Lambert_2024}. Even if we did not model the emission from the stellar population, it should contribute to relatively reddened optical distribution from positively shifted objects. Nevertheless, our color-magnitude estimate cannot be compared directly to observations, as we did not account for instrumental effects. 

Our results implicitly predict a non-negligible gamma-ray (and very-high-energy gamma-ray) emission from jet-star interactions. In fact, jet-star interactions are often listed to explain multi-wavelength, and especially gamma-ray emission in jetted AGN \citep{Barkov_2010, Bosch_2012, Delacita_2016, Vieyro_2017, Torres_2019}. This could explain, for example, and at least partly, extended emission detected from Centaurus\,A in the tera-electronvolt (TeV) band \citep{HESS_2020}, where the mass-loading scenario has been invoked to explain its broadband emission and the production of very-energetic cosmic rays \citep{Wykes_2013, Wykes_2014}. Accounting in the future for radiative cooling (if relevant) and the presence of other dissipative processes (as mass-entrainment at the jet boundaries) will refine our results, and assert the role of jet-star interactions in more complex jets. 

\subsection{Implications on the jet power}
\label{subsec:implications on the jet power}

 We have fixed the jet power at $L_{\rm j} = 10^{43}~\rm{erg}\cdot\rm{s}^{-1}$, a typical value for FR\,I jets \citep[e.g.,][]{Perucho_2020}. However, jet powers typically range from $L_{\rm j} = 10^{42}~\rm{erg}\cdot\rm{s}^{-1}$ to $L_{\rm j} \leq 10^{48}~\rm{erg}\cdot\rm{s}^{-1}$ \citep{Ghisellini_2001}. High jet powers reduce the effects of mass entrainment and energy dissipation due to mass loading \citep{Perucho_2014}, probably bringing the offset to zero or negative values. Conversely, lower jet powers efficiently dissipate energy along the deceleration region, which could explain the radio-optical offsets observed. For jets with powers higher than our initial value, detectable radio-optical offsets should persist down to the optical detection limit and turn out null or negative due to the lack of energy dissipation. Also in the case of lower jet powers, it should result in reduced offsets, converging to zero, due to intense energy dissipation.

In summary, under our framework there are three power regimes: 1) jets with the lowest powers would be decelerated close to the forming region, thus dissipating most of their kinetic energy, and this would result in zero or undetectable offsets; 2) The regime studied here would show positive offsets for the expected stellar populations in their host galaxies, and 3) high power jets undergo little dissipation within their host galaxies, as shown by their bulk velocities at kiloparsec scales, which means that they would tend to have zero or negative offsets, as expected from the conical jet expansion models \citep{Blandford_1979, Marscher_1985}. 

A basic question arises from the previous statements: why do \cite{Plavin_2019} observe such a fraction of positive offsets? The answer should be related to the jet-power distribution, as well as the stellar populations interacting with the jet. Too low power jets ($< 10^{42}~\rm{erg}\cdot\rm{s}^{-1}$) will not show extended emission, as they must be decelerated inside the host galaxy. Therefore, the fraction of positive offsets observed suggests a distribution of jet power peaking around $10^{43}~\rm{erg}\cdot\rm{s}^{-1}$. The presence of positive radio-optical shift as explained in our model will lead to low-power jets interacting with mainly main sequence stars with weak stellar winds within their first kiloparsecs (similar to the case studied here), and high-power jets with a lower density of red-giant stars. This aspect is explored in more detail in the next section.

\subsection{Implications on the mass-loading origin}
\label{subsec:imnplications on the stellar population}

All jets must entrain some amount of stellar matter, which may or may not significantly decelerate the jet, depending on the jet power and stellar populations \citep{Hubard_2006, Perucho_2014, Angles_2021}. If we assume that the mass entrained by the jet is only that driven by stellar winds, then the averaged mass-load necessary to account for the observed radio-optical offsets translate into an average stellar mass-loss rate per unit volume ranging from $10^{-10}$ to $10^{-9}~M_\odot \cdot \rm{yr}^{-1} \rm{pc}^{-3}$.

While K-M type stars are the most abundant in giant elliptical galaxies \citep{Dokkum_2010}, they have small mass-loss rates ($<~10^{-12}~M_\odot \cdot \rm{yr}^{-1}$) implying unrealistically large stellar densities to account for the needed mass loads. However, in the case of red giants, whose wind losses span from $10^{-10}$ to $10^{-5}~M_\odot \cdot \rm{yr}^{-1}$ \citep{Reimer_1975}, number densities of $10^{-3} - 1$ star per cubic parsec should be sufficient to account for the observed radio-optical offsets. The density of red-giants depends on the star formation history, and on the age of the galaxy. As jet-hosting galaxies have typically low star-forming rates and host old population of stars \citep{Zhu_2010, Heckman_2014}, it is safe to assume that red-giants are relatively abundant and their density within the range mentioned above. It is worth mentioning that jet-red-giant interactions has been already studied in the past to explain several aspects of jet physics and its potential impact on the non-thermal emission of jets \citep{Barkov_2010, Bosch_2012, Khangulyan_2013, Perucho_2017}. 

Our results suggest that misaligned jets, as those in radio galaxies and particularly Seyfert galaxies, should show a higher mass-load rate than for quasars in average, for fixed jet power, to explain the observed offsets. In the case where stellar wind dominates mass-load, this would require the presence of star forming regions, and different stellar populations with a larger fraction of very luminous stars that exhibits very high stellar wind loss, as we discuss below.

However, even if we assumed that the main source of dissipation is the interaction with stars, distinguishing between different average stellar mass-loss rates can be challenging due to overlapping effects, particularly at lower $d_{\rm app}$ values. This overlap introduces a degeneracy not only dependent on $r_{\rm c, s}$, the stellar core radius, but also on other factors such as the ambient gas and the jet properties. The observed degeneracy occurs because different combinations of mass-loading values and core radii can produce similar $d_{\rm app}$ outcomes. This is especially the case when only considering one observational parameter: For example, a larger gas core radius might mimic the effects of a higher mass-loading rate in terms of reducing $d_{\rm app}$.

The impact of other jet parameters and the power-law indices for ambient gas and stellar distributions on the radio-optical offset is complex and beyond the scope of this paper. However, one possible approach to address the $r_{\rm c, s}$ degeneracy is to analyze the average ratio between the radio and optical fluxes (estimated at $4.3 \times 10^{10}~\rm{Hz}$ and $5 \times 10^{14}~\rm{Hz}$, respectively) for all selected sources at a given average mass-load value $Q_0$. By averaging the flux ratio over many sources, the specific geometric conditions of individual sources tend to cancel out. The results are displayed in Figure~\ref{fig:ratio_opt_rad}. The figure shows that the differences in flux ratio between the two stellar core radii are limited but could allow us, through observational data, to distinguish the profile of mass-load. If the shift is mainly caused by stellar winds, this analysis would be feasible only for measured, positive radio-optical offsets. For instance, knowing $z$, the stellar mass-loss rates per unit volume (by characterizing the stellar populations and densities in galaxies) and $r_{\rm c, s}$ might help to constrain $\theta_{\rm obs}$. Alternatively, using independent measures of the viewing angle and $r_{\rm c, s}$, one could constrain the population and distribution of stars in the core of galaxies.

In high-power jets, the fraction of energy dissipated is much smaller and, although an increased average stellar mass-loss rate could result in observable radio-optical offsets (see Figure~\ref{fig:range_q0}), the values necessary to produce observed positive offsets are probably too large to be realistic. Therefore, the offsets are expected to be visible mainly in the case of low-power jets. In this case, we could expect observable offsets to emerge for more realistic average stellar mass-loss rates (see Figure~\ref{fig:evolv_rcs}). In all cases, though, a non-uniform stellar distribution, where densities vary significantly across the galactic atmosphere \citep{Gebhardt_2009, Vieyro_2017}, must be considered for quantitative comparisons with observations.

In a broader view, a more plausible hypothesis might be that mass-load could be driven not only by stellar wind but also by mass-entrainment. For instance, \cite{Matsumoto_2017, Gourgoliatos_2018,Perucho_2020} discussed how small-scale instabilities or the penetration of stars into the jet can initiate mixing layers, leading to mass-load and jet deceleration. This, of course, mitigates the capacity of our model to constraint the stellar population.

\subsection{Observational biases}
\label{subsec:observational biases}

Doppler boosting has a crucial impact on the detection of radio-optical offsets. We have shown that the sources showing an offset close to zero are over-represented in the set selected according to \emph{Gaia} sensitivity (see Figure~\ref{fig:histo_offset}), in agreement with \citet{Petrov_2017l, Petrov_2019}. This is mainly due to the contribution of jets with small viewing angles, for which the emission is Doppler-boosted. In that case, the majority of the sources are detected in the optical band regardless of the average stellar mass-loss rate (see Figure~\ref{fig:d_app_evolv_z}, right). 

The influence of cosmological distance in the detection is most evident in the case of low average mass-load ($Q_0 \leq 5 \times 10^{23}~\rm{g}\cdot\rm{yr}^{-1}\cdot\rm{pc}^{-3}$), where low-redshift sources can be over-represented (see Figure~\ref{fig:d_app_evolv_z}, left). This effect is weaker in the case of medium to high average mass-load rates ($Q_0 > 5 \times 10^{23}~\rm{g}\cdot\rm{yr}^{-1}\rm{pc}^{-3}$) because the strong dissipation makes them brighter and therefore detectable at larger distances. Thus, the recent and future improvements in \emph{Gaia}'s sensitivity, which has been enhanced through continuous advancements in its instruments and data processing techniques \citep{Gaia_2021}, are crucial to leveraging the impact of observational biases to fully test our predictions with observations. 

Finally, observational results from \cite{Secrest_2022} indicate that the prevalence of radio-optical offsets tends to decrease with increasing optical variability, which can be the consequence of the small line-of-sight angles typical of highly variable sources. This result will be studied in a follow-up research.

\subsection{Predictions from the mass-loading scenario}
\label{subsec:prediction from the mass-loading scenario}

The mass-loading scenario presented in this paper may provide not only predictions on measurable observables, such as radio-optical offsets and color magnitudes, but also constraints on the jet power, a parameter notoriously difficult to deduce from observations \citep{Hardcastle_2018, perucho_2019}. Jet power is typically inferred from secondary indicators like luminosity \citep{Daly_2012, Godfrey_2013} and emission line features, which can be affected by multiple factors, including Doppler boosting, viewing angle, and interaction with the ambient medium \citep{Ghisellini_2001, Fujita_2016, Foschini_2024}.

Our results trace a plausible, independent way to estimate jet powers in those cases in which radio-optical offsets are detected, which we explain here. For a given jet host galaxy with a known redshift $z$, optical observations can provide an estimate on the stellar distribution $r_{\rm c,s}$, the color index or the radio to optical flux ratio (see Figure~\ref{fig:ratio_opt_rad}), and together with radio observations, the radio-optical offset. 

The jet parameters that determine the observed radio-optical offset are the viewing angle $\theta_{\rm obs}$ and the jet power $L_{\rm j}$. If the former can be estimated by, e.g., jet-to-counter-jet flux ratio, our model can give an independent estimate of the jet power. In a lesser extent, depending on the fraction of mass-load driven by stellar winds, one could also derive information on the stellar types, and on the potential fraction of red-giants present in order to explain the observed radio-optical offset.

The jet power is determined by the kinetic, internal, and magnetic energy fluxes, so for a given value of power, different configurations are possible. Therefore, once power is estimated, and following \cite{Angles_2021}, we could further constrain the jet physics by defining regions of the parameter space for rest mass density, pressure, velocity, field intensity, etc. These different configurations can also influence the observed offset: in this paper, we have used a case where the jet is carrying a significant amount of kinetic energy from \cite{Angles_2021} as a base configuration, because these can efficiently dissipate energy and thus explain the observed offsets. A kinetically dominated jet appears to be not necessary to explain radio-optical offsets, a small amount of kinetic energy is dissipated. We have verified that internal dominated jet cases lead to null or negative offsets, and were excluded from this study according to our purposes. The comparisons of our predictions with observations will test the capacity of radio-optical offsets to open a new way to predict fundamental AGN properties \citep{Petrov_2017, Wang_2022}.

Future work should include the direct application of this method on specific sources and shed light on the validity of this scenario, utilizing detailed source analysis and setup to feed multidimensional, dynamical simulations.


\section{Conclusions}
\label{sec:Conclusions}

Jet-star interactions are crucial for understanding multi-wavelength observations and dynamics of jets, particularly in the context of FR I galaxies. The mass-loading scenario provides a promising framework for explaining partly the dynamics of kiloparsec-scale jets, especially when considering typical FR I jet power levels. In this study, we conducted RMHD simulations of mass-loaded stationary jets and used the radiative transfer code \texttt{RIPTIDE} to generate multi-wavelength synthetic synchrotron emission maps. Focusing on the galaxy properties, we thoroughly examined the radio and optical emission maps and the color magnitude. Our key findings are summarized as follows:

\begin{itemize}
    \item[$\bullet$] \emph{Jet deceleration and energy dissipation}: Jets with a power of $L_{\rm j} = 10^{43}~\rm{erg} \cdot \rm{s}^{-1}$ experience deceleration and convert kinetic energy into internal energy, accelerating particles in the jet. This dissipation leads to an increase of the minimal Lorentz factor $\gamma^\prime_{\rm e, min}$ of non-thermal electrons along the jet under the prescription chosen here. This behavior is observed in jets with a sufficient kinetic energy budget, as expected to be the case at the beginning of the deceleration/dissipation region \citep{Laing_2014}. \\
    
    \item[$\bullet$] \emph{Radio-optical offsets}: The increase of $\gamma^\prime_{\rm e, min}$ results in spatially separated radio and optical peak emission regions. The radio-dominated region is located closer to the jet base where $\gamma^\prime_{\rm e, min}$ is low, while the optical-dominated region is located downstream close to the position where dissipation occurs, causing a positive radio-optical offset. \\
    
    \item[$\bullet$] \emph{Impact of galaxy properties}: The distribution of gas and stars in the host galaxy significantly influences the observed offset. Gas distribution, represented by its core size, $r_{\rm c}$, impacts mainly the location of the radio-dominated region, while the stellar distribution, represented by $r_{\rm c, s}$, changes the position where most of the dissipation occurs. There is a fine balance between these two parameters, which results either in an extension or a limitation of the radio-optical offset. \\

    \item[$\bullet$] \emph{Useful observables}: The radio-to-optical flux ratio and the color magnitude permit us to derive the average stellar mass-loss rate present in the jet. We conclude that lower flux ratios and higher optical emission from the jet are linked to lower average stellar mass-loss rates and result in a more significant (positive) radio-optical offset. \\
    
    \item[$\bullet$] \emph{Observational biases}: The viewing angle of the jet and redshift affect the observed radio-optical offsets. Offset detectability is also intimately linked to flux sensitivity, especially in the optical band. We expect that low redshift and small viewing angles are over-represented in any dataset. \\

    \item[$\bullet$] \emph{Observational evidence}: Our work aligns with collected evidence that energy dissipation, therefore non-thermal emission, occurs in the first few jet kiloparsecs \citep{Laing_2014}, also supported by optical polarimetry \citep{Kovalev_2020}. For a certain average mass-load rate regime, we were able to reproduce the typical range of radio-optical offset properties observed, and thus for different types of AGN jets according to their viewing angle. In the case where mass-load is mainly driven by stellar winds, our results point at a dominant population of red-giants interacting with the jet, with a plausible range of number density in the galactic core. The trend of the color magnitude of our simulated jets concerning the radio-optical offset is consistent with observations, even if we are not accounting for optical emission from a stellar population, as its impact should be limited.  \\

    \item[$\bullet$] \emph{Observational predictions}: From this scenario, and our results, we can predict that radio-optical offsets should evolve accordingly to the jet power at play. Low power jets ($L_{\rm j} < 10^{43}~\rm{erg}\cdot\rm{s}^{-1}$), should display null or negative offsets, with a more stable optical emission. Larger power FRI jets ($ 10^{43}~\rm{erg}\cdot\rm{s}^{-1} < L_{\rm j} < 10^{44}~\rm{erg}\cdot\rm{s}^{-1}$) should display observable, positive offsets. FR\,II jets ($L_{\rm j} > 10^{44}~\rm{erg}\cdot\rm{s}^{-1}$) should display null or negative offsets as the jet evolves in a conical way. For a given jet power, increasing (decreasing) the average mass-load rate should decrease (increase) the apparent radio-optical offset. If we only consider stellar winds, this could be explained by different types of stellar populations dominating the average mass-loss rates. Mass-entrainment from jet-star interactions could also imply a significant amount of mass-load in the jet, and is known to influence the jet behaviors and non-thermal emission.
    
    To a lesser extent, we predict the presence of an extended gamma-ray emission component caused by inverse Compton, and the production of very energetic cosmic rays (protons, neutrinos) from mass-loaded jets. Notably in the case of jet-red-giant interactions, local jet-star interactions could produce high and very-high-energy gamma-ray emission \citep{Torres_2019}, joined with a production of high-energy cosmic rays as neutrinos \citep{Wykes_2014, Wykes_2018, Wang_2022}.
\end{itemize}

Future work will focus on the study of the interplay between jet power and the average mass-load rate. This will come with incorporation of more detailed observational constraints from optical astronomy (as \emph{Gaia}'s angular resolution). 
A dedicated application on specific sources will be done by fixing a range of parameters as the radio-optical offset, from observations. This may allow us to understand the origin of the mass-load derived here, and potentially give an independent estimate of the jet power in those cases, and study the jet physics at a unique level, offering a new multiwavelength view of the AGN classification.  


\begin{acknowledgements}
We thank the reviewer for the constructive comments that improved the quality of the present manuscript. We thank Andrei Lobanov for valuable comments on the manuscript. This work has been supported by the Spanish Ministry of Science through Grant \texttt{PID2022-136828NB-C43}, from the Generalitat Valenciana through grant \texttt{CIPROM/2022/49}, from the Astrophysics and High Energy Physics program supported by the Spanish Ministry of Science and Generalitat Valenciana with funding from European Union NextGenerationEU (\texttt{PRTR-C17.I1}) through grant \texttt{ASFAE/2022/005}.
YYK was supported by the MuSES project, which has received funding from the European Research Council (ERC) under the European Union’s Horizon 2020 Research and Innovation Programme (grant agreement No 101142396).
\end{acknowledgements}


\bibliographystyle{aa.bst} 
\bibliography{bib.bib} 

\begin{thebibliography}{80}
\expandafter\ifx\csname natexlab\endcsname\relax\def\natexlab#1{#1}\fi

\bibitem[{{Angl{\'e}s-Castillo} {et~al.}(2021){Angl{\'e}s-Castillo}, {Perucho},
  {Mart{\'\i}}, \& {Laing}}]{Angles_2021}
{Angl{\'e}s-Castillo}, A., {Perucho}, M., {Mart{\'\i}}, J.~M., \& {Laing},
  R.~A. 2021, MNRAS, 500, 1512

\bibitem[{{Barkov} {et~al.}(2010){Barkov}, {Aharonian}, \&
  {Bosch-Ramon}}]{Barkov_2010}
{Barkov}, M.~V., {Aharonian}, F.~A., \& {Bosch-Ramon}, V. 2010, \apj, 724, 1517

\bibitem[{{Bicknell}(1984)}]{Bicknell_1984}
{Bicknell}, G.~V. 1984, \apj, 286, 68

\bibitem[{{Bicknell}(1994)}]{Bicknell_1994}
{Bicknell}, G.~V. 1994, \apj, 422, 542

\bibitem[{{Blandford} {et~al.}(2019){Blandford}, {Meier}, \&
  {Readhead}}]{Blandford_2019}
{Blandford}, R., {Meier}, D., \& {Readhead}, A. 2019, \araa, 57, 467

\bibitem[{{Blandford} \& {K{\"o}nigl}(1979)}]{Blandford_1979}
{Blandford}, R.~D. \& {K{\"o}nigl}, A. 1979, \apj, 232, 34

\bibitem[{{Boccardi} {et~al.}(2017){Boccardi}, {Krichbaum}, {Ros}, \&
  {Zensus}}]{Boccardi_2017}
{Boccardi}, B., {Krichbaum}, T.~P., {Ros}, E., \& {Zensus}, J.~A. 2017, \aapr,
  25, 4

\bibitem[{{Bosch-Ramon} {et~al.}(2012){Bosch-Ramon}, {Perucho}, \&
  {Barkov}}]{Bosch_2012}
{Bosch-Ramon}, V., {Perucho}, M., \& {Barkov}, M.~V. 2012, \aap, 539, A69

\bibitem[{{B{\"o}ttcher} \& {Dermer}(2010)}]{Bottcher_2010}
{B{\"o}ttcher}, M. \& {Dermer}, C.~D. 2010, \apj, 711, 445

\bibitem[{{Bowman} {et~al.}(1996){Bowman}, {Leahy}, \&
  {Komissarov}}]{Bowman_1996}
{Bowman}, M., {Leahy}, J.~P., \& {Komissarov}, S.~S. 1996, \mnras, 279, 899

\bibitem[{{Bridle} \& {Perley}(1984)}]{Briddle_1984}
{Bridle}, A.~H. \& {Perley}, R.~A. 1984, \araa, 22, 319

\bibitem[{{Choi} \& {Wiita}(2010)}]{Choi_2010}
{Choi}, E. \& {Wiita}, P.~J. 2010, \apjs, 191, 113

\bibitem[{{Daly} {et~al.}(2012){Daly}, {Sprinkle}, {O'Dea}, {Kharb}, \&
  {Baum}}]{Daly_2012}
{Daly}, R.~A., {Sprinkle}, T.~B., {O'Dea}, C.~P., {Kharb}, P., \& {Baum}, S.~A.
  2012, \mnras, 423, 2498

\bibitem[{{de la Cita} {et~al.}(2016){de la Cita}, {Bosch-Ramon},
  {Paredes-Fortuny}, {Khangulyan}, \& {Perucho}}]{Delacita_2016}
{de la Cita}, V.~M., {Bosch-Ramon}, V., {Paredes-Fortuny}, X., {Khangulyan},
  D., \& {Perucho}, M. 2016, \aap, 591, A15

\bibitem[{{Falle}(1991)}]{Falle_1991}
{Falle}, S.~A.~E.~G. 1991, \mnras, 250, 581

\bibitem[{{Fanaroff} \& {Riley}(1974)}]{Fanaroff_1974}
{Fanaroff}, B.~L. \& {Riley}, J.~M. 1974, MNRAS, 167, 31P

\bibitem[{{Fichet de Clairfontaine} {et~al.}(2022){Fichet de Clairfontaine},
  {Meliani}, \& {Zech}}]{Fichet_2022}
{Fichet de Clairfontaine}, G., {Meliani}, Z., \& {Zech}, A. 2022, \aap, 661,
  A54

\bibitem[{{Fichet de Clairfontaine} {et~al.}(2021){Fichet de Clairfontaine},
  {Meliani}, {Zech}, \& {Hervet}}]{Fichet_2021}
{Fichet de Clairfontaine}, G., {Meliani}, Z., {Zech}, A., \& {Hervet}, O. 2021,
  \aap, 647, A77

\bibitem[{Foschini {et~al.}(2024)Foschini, Dalla~Barba, Tornikoski, Andernach,
  Marziani, Marscher, Jorstad, Järvelä, Antón, \&
  Dalla~Bontà}]{Foschini_2024}
Foschini, L., Dalla~Barba, B., Tornikoski, M., {et~al.} 2024, Universe, 10

\bibitem[{{Fromm} {et~al.}(2016){Fromm}, {Perucho}, {Mimica}, \&
  {Ros}}]{Fromm_2016}
{Fromm}, C.~M., {Perucho}, M., {Mimica}, P., \& {Ros}, E. 2016, \aap, 588, A101

\bibitem[{{Fujita} {et~al.}(2016){Fujita}, {Kawakatu}, \&
  {Shlosman}}]{Fujita_2016}
{Fujita}, Y., {Kawakatu}, N., \& {Shlosman}, I. 2016, \pasj, 68, 26

\bibitem[{{Gaia Collaboration} {et~al.}(2021){Gaia Collaboration}, {Brown},
  {Vallenari}, {Prusti}, {de Bruijne}, {Babusiaux}, {Biermann}, {Creevey},
  {Evans}, {Eyer}, {Hutton}, {Jansen}, {Jordi}, {Klioner}, {Lammers},
  {Lindegren}, {Luri}, {Mignard}, {Panem}, {Pourbaix}, {Randich}, {Sartoretti},
  {Soubiran}, {Walton}, {Arenou}, {Bailer-Jones}, {Bastian}, {Cropper},
  {Drimmel}, {Katz}, {Lattanzi}, {van Leeuwen}, {Bakker}, {Cacciari},
  {Casta{\~n}eda}, {De Angeli}, {Ducourant}, {Fabricius}, {Fouesneau},
  {Fr{\'e}mat}, {Guerra}, {Guerrier}, {Guiraud}, {Jean-Antoine Piccolo},
  {Masana}, {Messineo}, {Mowlavi}, {Nicolas}, {Nienartowicz}, {Pailler},
  {Panuzzo}, {Riclet}, {Roux}, {Seabroke}, {Sordo}, {Tanga}, {Th{\'e}venin},
  {Gracia-Abril}, {Portell}, {Teyssier}, {Altmann}, {Andrae}, {Bellas-Velidis},
  {Benson}, {Berthier}, {Blomme}, {Brugaletta}, {Burgess}, {Busso}, {Carry},
  {Cellino}, {Cheek}, {Clementini}, {Damerdji}, {Davidson}, {Delchambre},
  {Dell'Oro}, {Fern{\'a}ndez-Hern{\'a}ndez}, {Galluccio}, {Garc{\'\i}a-Lario},
  {Garcia-Reinaldos}, {Gonz{\'a}lez-N{\'u}{\~n}ez}, {Gosset}, {Haigron},
  {Halbwachs}, {Hambly}, {Harrison}, {Hatzidimitriou}, {Heiter},
  {Hern{\'a}ndez}, {Hestroffer}, {Hodgkin}, {Holl}, {Jan{\ss}en}, {Jevardat de
  Fombelle}, {Jordan}, {Krone-Martins}, {Lanzafame}, {L{\"o}ffler}, {Lorca},
  {Manteiga}, {Marchal}, {Marrese}, {Moitinho}, {Mora}, {Muinonen}, {Osborne},
  {Pancino}, {Pauwels}, {Petit}, {Recio-Blanco}, {Richards}, {Riello},
  {Rimoldini}, {Robin}, {Roegiers}, {Rybizki}, {Sarro}, {Siopis}, {Smith},
  {Sozzetti}, {Ulla}, {Utrilla}, {van Leeuwen}, {van Reeven}, {Abbas}, {Abreu
  Aramburu}, {Accart}, {Aerts}, {Aguado}, {Ajaj}, {Altavilla}, {{\'A}lvarez},
  {{\'A}lvarez Cid-Fuentes}, {Alves}, {Anderson}, {Anglada Varela}, {Antoja},
  {Audard}, {Baines}, {Baker}, {Balaguer-N{\'u}{\~n}ez}, {Balbinot}, {Balog},
  {Barache}, {Barbato}, {Barros}, {Barstow}, {Bartolom{\'e}}, {Bassilana},
  {Bauchet}, {Baudesson-Stella}, {Becciani}, {Bellazzini}, {Bernet}, {Bertone},
  {Bianchi}, {Blanco-Cuaresma}, {Boch}, {Bombrun}, {Bossini}, {Bouquillon},
  {Bragaglia}, {Bramante}, {Breedt}, {Bressan}, {Brouillet}, {Bucciarelli},
  {Burlacu}, {Busonero}, {Butkevich}, {Buzzi}, {Caffau}, {Cancelliere},
  {C{\'a}novas}, {Cantat-Gaudin}, {Carballo}, {Carlucci}, {Carnerero},
  {Carrasco}, {Casamiquela}, {Castellani}, {Castro-Ginard}, {Castro Sampol},
  {Chaoul}, {Charlot}, {Chemin}, {Chiavassa}, {Cioni}, {Comoretto}, {Cooper},
  {Cornez}, {Cowell}, {Crifo}, {Crosta}, {Crowley}, {Dafonte}, {Dapergolas},
  {David}, {David}, {de Laverny}, {De Luise}, {De March}, {De Ridder}, {de
  Souza}, {de Teodoro}, {de Torres}, {del Peloso}, {del Pozo}, {Delbo},
  {Delgado}, {Delgado}, {Delisle}, {Di Matteo}, {Diakite}, {Diener},
  {Distefano}, {Dolding}, {Eappachen}, {Edvardsson}, {Enke}, {Esquej}, {Fabre},
  {Fabrizio}, {Faigler}, {Fedorets}, {Fernique}, {Fienga}, {Figueras},
  {Fouron}, {Fragkoudi}, {Fraile}, {Franke}, {Gai}, {Garabato},
  {Garcia-Gutierrez}, {Garc{\'\i}a-Torres}, {Garofalo}, {Gavras}, {Gerlach},
  {Geyer}, {Giacobbe}, {Gilmore}, {Girona}, {Giuffrida}, {Gomel}, {Gomez},
  {Gonzalez-Santamaria}, {Gonz{\'a}lez-Vidal}, {Granvik},
  {Guti{\'e}rrez-S{\'a}nchez}, {Guy}, {Hauser}, {Haywood}, {Helmi}, {Hidalgo},
  {Hilger}, {H{\l}adczuk}, {Hobbs}, {Holland}, {Huckle}, {Jasniewicz},
  {Jonker}, {Juaristi Campillo}, {Julbe}, {Karbevska}, {Kervella}, {Khanna},
  {Kochoska}, {Kontizas}, {Kordopatis}, {Korn}, {Kostrzewa-Rutkowska},
  {Kruszy{\'n}ska}, {Lambert}, {Lanza}, {Lasne}, {Le Campion}, {Le Fustec},
  {Lebreton}, {Lebzelter}, {Leccia}, {Leclerc}, {Lecoeur-Taibi}, {Liao},
  {Licata}, {Lindstr{\o}m}, {Lister}, {Livanou}, {Lobel}, {Madrero Pardo},
  {Managau}, {Mann}, {Marchant}, {Marconi}, {Marcos Santos}, {Marinoni},
  {Marocco}, {Marshall}, {Martin Polo}, {Mart{\'\i}n-Fleitas}, {Masip},
  {Massari}, {Mastrobuono-Battisti}, {Mazeh}, {McMillan}, {Messina},
  {Michalik}, {Millar}, {Mints}, {Molina}, {Molinaro}, {Moln{\'a}r},
  {Montegriffo}, {Mor}, {Morbidelli}, {Morel}, {Morris}, {Mulone}, {Munoz},
  {Muraveva}, {Murphy}, {Musella}, {Noval}, {Ord{\'e}novic}, {Orr{\`u}},
  {Osinde}, {Pagani}, {Pagano}, {Palaversa}, {Palicio}, {Panahi}, {Pawlak},
  {Pe{\~n}alosa Esteller}, {Penttil{\"a}}, {Piersimoni}, {Pineau}, {Plachy},
  {Plum}, {Poggio}, {Poretti}, {Poujoulet}, {Pr{\v{s}}a}, {Pulone}, {Racero},
  {Ragaini}, {Rainer}, {Raiteri}, {Rambaux}, {Ramos}, {Ramos-Lerate}, {Re
  Fiorentin}, {Regibo}, {Reyl{\'e}}, {Ripepi}, {Riva}, {Rixon}, {Robichon},
  {Robin}, {Roelens}, {Rohrbasser}, {Romero-G{\'o}mez}, {Rowell}, {Royer},
  {Rybicki}, {Sadowski}, {Sagrist{\`a} Sell{\'e}s}, {Sahlmann}, {Salgado},
  {Salguero}, {Samaras}, {Sanchez Gimenez}, {Sanna}, {Santove{\~n}a},
  {Sarasso}, {Schultheis}, {Sciacca}, {Segol}, {Segovia}, {S{\'e}gransan},
  {Semeux}, {Shahaf}, {Siddiqui}, {Siebert}, {Siltala}, {Slezak}, {Smart},
  {Solano}, {Solitro}, {Souami}, {Souchay}, {Spagna}, {Spoto}, {Steele},
  {Steidelm{\"u}ller}, {Stephenson}, {S{\"u}veges}, {Szabados}, {Szegedi-Elek},
  {Taris}, {Tauran}, {Taylor}, {Teixeira}, {Thuillot}, {Tonello}, {Torra},
  {Torra}, {Turon}, {Unger}, {Vaillant}, {van Dillen}, {Vanel}, {Vecchiato},
  {Viala}, {Vicente}, {Voutsinas}, {Weiler}, {Wevers}, {Wyrzykowski}, {Yoldas},
  {Yvard}, {Zhao}, {Zorec}, {Zucker}, {Zurbach}, \& {Zwitter}}]{Gaia_2021}
{Gaia Collaboration}, {Brown}, A.~G.~A., {Vallenari}, A., {et~al.} 2021, \aap,
  649, A1

\bibitem[{{Gebhardt} \& {Thomas}(2009)}]{Gebhardt_2009}
{Gebhardt}, K. \& {Thomas}, J. 2009, \apj, 700, 1690

\bibitem[{{Ghisellini} \& {Celotti}(2001)}]{Ghisellini_2001}
{Ghisellini}, G. \& {Celotti}, A. 2001, \aap, 379, L1

\bibitem[{Godfrey \& Shabala(2013)}]{Godfrey_2013}
Godfrey, L. E.~H. \& Shabala, S.~S. 2013, The Astrophysical Journal, 767, 12

\bibitem[{{Gourgouliatos} \& {Komissarov}(2018)}]{Gourgoliatos_2018}
{Gourgouliatos}, K.~N. \& {Komissarov}, S.~S. 2018, Nature Astronomy, 2, 167

\bibitem[{{Gómez} {et~al.}(1995){Gómez}, {Martí}, {Marscher}, {Ibánez}, \&
  {Marcaide}}]{Gomez_1995}
{Gómez}, J.~L., {Martí}, J.~M.~A., {Marscher}, A.~P., {Ibánez}, J.~M.~A., \&
  {Marcaide}, J.~M. 1995, \apjl, 449, L19

\bibitem[{{H.~E.~S.~S. Collaboration} {et~al.}(2020){H.~E.~S.~S.
  Collaboration}, {Abdalla}, {Adam}, {Aharonian}, {Ait Benkhali},
  {Ang{\"u}ner}, {Arakawa}, {Arcaro}, {Armand}, {Ashkar}, {Backes}, {Barbosa
  Martins}, {Barnard}, {Becherini}, {Berge}, {Bernl{\"o}hr}, {Blackwell},
  {B{\"o}ttcher}, {Boisson}, {Bolmont}, {Bonnefoy}, {Bregeon}, {Breuhaus},
  {Brun}, {Brun}, {Bryan}, {B{\"u}chele}, {Bulik}, {Bylund}, {Capasso},
  {Caroff}, {Carosi}, {Casanova}, {Cerruti}, {Chand}, {Chandra}, {Chen},
  {Colafrancesco}, {Cury{\l}o}, {Davids}, {Deil}, {Devin}, {deWilt}, {Dirson},
  {Djannati-Ata{\"\i}}, {Dmytriiev}, {Donath}, {Doroshenko}, {Drury}, {Dyks},
  {Egberts}, {Emery}, {Ernenwein}, {Eschbach}, {Feijen}, {Fegan}, {Fiasson},
  {Fontaine}, {Funk}, {F{\"u}{\ss}ling}, {Gabici}, {Gallant}, {Gat{\'e}},
  {Giavitto}, {Glawion}, {Glicenstein}, {Gottschall}, {Grondin}, {Hahn},
  {Haupt}, {Heinzelmann}, {Henri}, {Hermann}, {Hinton}, {Hofmann}, {Hoischen},
  {Holch}, {Holler}, {Horns}, {Huber}, {Iwasaki}, {Jamrozy}, {Jankowsky},
  {Jankowsky}, {Jardin-Blicq}, {Jung-Richardt}, {Kastendieck},
  {Katarzy{\'n}ski}, {Katsuragawa}, {Katz}, {Khangulyan}, {Kh{\'e}lifi},
  {King}, {Klepser}, {Klu{\'z}niak}, {Komin}, {Kosack}, {Kostunin}, {Kraus},
  {Lamanna}, {Lau}, {Lemi{\`e}re}, {Lemoine-Goumard}, {Lenain}, {Leser},
  {Levy}, {Lohse}, {Lypova}, {Mackey}, {Majumdar}, {Malyshev}, {Marandon},
  {Marcowith}, {Mares}, {Mariaud}, {Mart{\'\i}-Devesa}, {Marx}, {Maurin},
  {Meintjes}, {Mitchell}, {Moderski}, {Mohamed}, {Mohrmann}, {Moore}, {Moulin},
  {Muller}, {Murach}, {Nakashima}, {de Naurois}, {Ndiyavala}, {Niederwanger},
  {Niemiec}, {Oakes}, {O'Brien}, {Odaka}, {Ohm}, {de Ona Wilhelmi},
  {Ostrowski}, {Oya}, {Panter}, {Parsons}, {Perennes}, {Petrucci}, {Peyaud},
  {Piel}, {Pita}, {Poireau}, {Priyana Noel}, {Prokhorov}, {Prokoph},
  {P{\"u}hlhofer}, {Punch}, {Quirrenbach}, {Raab}, {Rauth}, {Reimer}, {Reimer},
  {Remy}, {Renaud}, {Rieger}, {Rinchiuso}, {Romoli}, {Rowell}, {Rudak},
  {Ruiz-Velasco}, {Sahakian}, {Saito}, {Sanchez}, {Santangelo}, {Sasaki},
  {Schlickeiser}, {Sch{\"u}ssler}, {Schulz}, {Schutte}, {Schwanke},
  {Schwemmer}, {Seglar-Arroyo}, {Senniappan}, {Seyffert}, {Shafi},
  {Shiningayamwe}, {Simoni}, {Sinha}, {Sol}, {Specovius}, {Spir-Jacob},
  {Stawarz}, {Steenkamp}, {Stegmann}, {Steppa}, {Takahashi}, {Tavernier},
  {Taylor}, {Terrier}, {Tiziani}, {Tluczykont}, {Trichard}, {Tsirou}, {Tsuji},
  {Tuffs}, {Uchiyama}, {van der Walt}, {van Eldik}, {van Rensburg}, {van
  Soelen}, {Vasileiadis}, {Veh}, {Venter}, {Vincent}, {Vink}, {Voisin},
  {V{\"o}lk}, {Vuillaume}, {Wadiasingh}, {Wagner}, {White}, {Wierzcholska},
  {Yang}, {Yoneda}, {Zacharias}, {Zanin}, {Zdziarski}, {Zech}, {Ziegler},
  {Zorn}, \& {{\.Z}ywucka}}]{HESS_2020}
{H.~E.~S.~S. Collaboration}, {Abdalla}, H., {Adam}, R., {et~al.} 2020, \nat,
  582, 356

\bibitem[{{Hardcastle}(2018)}]{Hardcastle_2018}
{Hardcastle}, M. 2018, Nature Astronomy, 2, 273

\bibitem[{{Heckman} \& {Best}(2014)}]{Heckman_2014}
{Heckman}, T.~M. \& {Best}, P.~N. 2014, \araa, 52, 589

\bibitem[{{Hubbard} \& {Blackman}(2006)}]{Hubard_2006}
{Hubbard}, A. \& {Blackman}, E.~G. 2006, MNRAS, 371, 1717

\bibitem[{{Jordi} {et~al.}(2010){Jordi}, {Gebran}, {Carrasco}, {de Bruijne},
  {Voss}, {Fabricius}, {Knude}, {Vallenari}, {Kohley}, \& {Mora}}]{Jordi_2010}
{Jordi}, C., {Gebran}, M., {Carrasco}, J.~M., {et~al.} 2010, \aap, 523, A48

\bibitem[{{Katarzy{\'n}ski} {et~al.}(2001){Katarzy{\'n}ski}, {Sol}, \&
  {Kus}}]{Katarzynski_2001}
{Katarzy{\'n}ski}, K., {Sol}, H., \& {Kus}, A. 2001, \aap, 367, 809

\bibitem[{{Khangulyan} {et~al.}(2013){Khangulyan}, {Barkov}, {Bosch-Ramon},
  {Aharonian}, \& {Dorodnitsyn}}]{Khangulyan_2013}
{Khangulyan}, D.~V., {Barkov}, M.~V., {Bosch-Ramon}, V., {Aharonian}, F.~A., \&
  {Dorodnitsyn}, A.~V. 2013, \apj, 774, 113

\bibitem[{{Komissarov}(1994)}]{Komissarov_1994}
{Komissarov}, S.~S. 1994, MNRAS, 269, 394

\bibitem[{Komissarov(2012)}]{Komissarov_2012}
Komissarov, S.~S. 2012, MNRAS, 422, 326

\bibitem[{{Komissarov} {et~al.}(2015){Komissarov}, {Porth}, \&
  {Lyutikov}}]{Komissarov_2015}
{Komissarov}, S.~S., {Porth}, O., \& {Lyutikov}, M. 2015, Computational
  Astrophysics and Cosmology, 2, 9

\bibitem[{{Komissarov} {et~al.}(2009){Komissarov}, {Vlahakis}, {K{\"o}nigl}, \&
  {Barkov}}]{Komissarov_2009}
{Komissarov}, S.~S., {Vlahakis}, N., {K{\"o}nigl}, A., \& {Barkov}, M.~V. 2009,
  \mnras, 394, 1182

\bibitem[{{Kovalev} {et~al.}(2017){Kovalev}, {Petrov}, \&
  {Plavin}}]{Kovalev_2017}
{Kovalev}, Y.~Y., {Petrov}, L., \& {Plavin}, A.~V. 2017, \aap, 598, L1

\bibitem[{Kovalev {et~al.}(2020)Kovalev, Zobnina, Plavin, \&
  Blinov}]{Kovalev_2020}
Kovalev, Y.~Y., Zobnina, D.~I., Plavin, A.~V., \& Blinov, D. 2020, MNRAS:
  Letters, 493, L54

\bibitem[{Laing {et~al.}(1996)Laing, Hardee, Bridle, \& Zensus}]{Laing_1996}
Laing, R., Hardee, P., Bridle, A., \& Zensus, J. 1996, in Astronomical Society
  of the Pacific Conference Series, Vol. 100, 241

\bibitem[{{Laing} \& {Bridle}(2014)}]{Laing_2014}
{Laing}, R.~A. \& {Bridle}, A.~H. 2014, MNRAS, 437, 3405

\bibitem[{{Lambert} {et~al.}(2024){Lambert}, {Sol}, \&
  {Pierron}}]{Lambert_2024}
{Lambert}, S., {Sol}, H., \& {Pierron}, A. 2024, \aap, 684, A202

\bibitem[{{Lemoine} \& {Pelletier}(2003)}]{Lemoine_2003}
{Lemoine}, M. \& {Pelletier}, G. 2003, \apjl, 589, L73

\bibitem[{{Marscher} \& {Gear}(1985)}]{Marscher_1985}
{Marscher}, A.~P. \& {Gear}, W.~K. 1985, The Astrophysical Journal, 298, 114

\bibitem[{{Mart{\'\i}}(2015)}]{Marti_2015}
{Mart{\'\i}}, J.~M. 2015, MNRAS, 452, 3106

\bibitem[{{Mart{\'\i}} {et~al.}(2016){Mart{\'\i}}, {Perucho}, \&
  {G{\'o}mez}}]{Marti_2016}
{Mart{\'\i}}, J.~M., {Perucho}, M., \& {G{\'o}mez}, J.~L. 2016, \apj, 831, 163

\bibitem[{{Matsumoto} {et~al.}(2017){Matsumoto}, {Aloy}, \&
  {Perucho}}]{Matsumoto_2017}
{Matsumoto}, J., {Aloy}, M.~A., \& {Perucho}, M. 2017, \mnras, 472, 1421

\bibitem[{{Matsumoto} \& {Masada}(2013)}]{Matsumoto_2013}
{Matsumoto}, J. \& {Masada}, Y. 2013, \apjl, 772, L1

\bibitem[{{Mimica} \& {Aloy}(2012)}]{Mimica_2012}
{Mimica}, P. \& {Aloy}, M.~A. 2012, \mnras, 421, 2635

\bibitem[{{Mimica} {et~al.}(2009){Mimica}, {Aloy}, {Agudo}, {Mart{\'\i}},
  {G{\'o}mez}, \& {Miralles}}]{Mimica_2009}
{Mimica}, P., {Aloy}, M.~A., {Agudo}, I., {et~al.} 2009, \apj, 696, 1142

\bibitem[{Mingo {et~al.}(2019)Mingo, Croston, Hardcastle, Best, Duncan,
  Morganti, Rottgering, Sabater, Shimwell, Williams, Brienza, Gurkan, Mahatma,
  Morabito, Prandoni, Bondi, Ineson, \& Mooney}]{Mingo_2019}
Mingo, B., Croston, J.~H., Hardcastle, M.~J., {et~al.} 2019, MNRAS, 488, 2701

\bibitem[{{Mizuno} {et~al.}(2015){Mizuno}, {G{\'o}mez}, {Nishikawa}, {Meli},
  {Hardee}, \& {Rezzolla}}]{Mizuno_2015}
{Mizuno}, Y., {G{\'o}mez}, J.~L., {Nishikawa}, K.-I., {et~al.} 2015, \apj, 809,
  38

\bibitem[{{Netzer}(2015)}]{Netzer_2015}
{Netzer}, H. 2015, \araa, 53, 365

\bibitem[{{Ostrowski} \& {Bednarz}(2002)}]{Ostrowski_2002}
{Ostrowski}, M. \& {Bednarz}, J. 2002, \aap, 394, 1141

\bibitem[{Perucho(2019)}]{perucho_2019}
Perucho, M. 2019, Galaxies, 7

\bibitem[{{Perucho}(2020)}]{Perucho_2020}
{Perucho}, M. 2020, MNRAS, 494, L22

\bibitem[{{Perucho} {et~al.}(2017){Perucho}, {Bosch-Ramon}, \&
  {Barkov}}]{Perucho_2017}
{Perucho}, M., {Bosch-Ramon}, V., \& {Barkov}, M.~V. 2017, \aap, 606, A40

\bibitem[{{Perucho} \& {Mart{\'\i}}(2007)}]{Perucho_2007}
{Perucho}, M. \& {Mart{\'\i}}, J.~M. 2007, MNRAS, 382, 526

\bibitem[{{Perucho} {et~al.}(2014){Perucho}, {Mart{\'\i}}, {Laing}, \&
  {Hardee}}]{Perucho_2014}
{Perucho}, M., {Mart{\'\i}}, J.~M., {Laing}, R.~A., \& {Hardee}, P.~E. 2014,
  MNRAS, 441, 1488

\bibitem[{{Petrov} \& {Kovalev}(2017{\natexlab{a}})}]{Petrov_2017}
{Petrov}, L. \& {Kovalev}, Y.~Y. 2017{\natexlab{a}}, MNRAS, 471, 3775

\bibitem[{{Petrov} \& {Kovalev}(2017{\natexlab{b}})}]{Petrov_2017l}
{Petrov}, L. \& {Kovalev}, Y.~Y. 2017{\natexlab{b}}, \mnras, 467, L71

\bibitem[{{Petrov} {et~al.}(2019){Petrov}, {Kovalev}, \&
  {Plavin}}]{Petrov_2019}
{Petrov}, L., {Kovalev}, Y.~Y., \& {Plavin}, A.~V. 2019, \mnras, 482, 3023

\bibitem[{{Plavin} {et~al.}(2019){Plavin}, {Kovalev}, \&
  {Petrov}}]{Plavin_2019}
{Plavin}, A.~V., {Kovalev}, Y.~Y., \& {Petrov}, L.~Y. 2019, \apj, 871, 143

\bibitem[{{Plavin} {et~al.}(2022){Plavin}, {Kovalev}, \&
  {Pushkarev}}]{Plavin_2022}
{Plavin}, A.~V., {Kovalev}, Y.~Y., \& {Pushkarev}, A.~B. 2022, \apjs, 260, 4

\bibitem[{{Reimers}(1975)}]{Reimer_1975}
{Reimers}, D. 1975, Memoires of the Societe Royale des Sciences de Liege, 8,
  369

\bibitem[{{Ricci} {et~al.}(2024){Ricci}, {Perucho}, {L{\'o}pez-Miralles},
  {Mart{\'\i}}, \& {Boccardi}}]{Ricci_2024}
{Ricci}, L., {Perucho}, M., {L{\'o}pez-Miralles}, J., {Mart{\'\i}}, J.~M., \&
  {Boccardi}, B. 2024, \aap, 683, A235

\bibitem[{{Rodriguez Espinosa} {et~al.}(1987){Rodriguez Espinosa}, {Rudy}, \&
  {Jones}}]{Rodriguez_1987}
{Rodriguez Espinosa}, J.~M., {Rudy}, R.~J., \& {Jones}, B. 1987, \apj, 312, 555

\bibitem[{{Rybicki} \& {Lightman}(1979)}]{Rybicki_1979}
{Rybicki}, G.~B. \& {Lightman}, A.~P. 1979, {Radiative processes in
  astrophysics}

\bibitem[{{Secrest}(2022)}]{Secrest_2022}
{Secrest}, N.~J. 2022, \apjl, 939, L32

\bibitem[{Synge \& Morse(1958)}]{Synge_1958}
Synge, J.~L. \& Morse, P.~M. 1958, The relativistic gas

\bibitem[{{Torres-Alb{\`a}} \& {Bosch-Ramon}(2019)}]{Torres_2019}
{Torres-Alb{\`a}}, N. \& {Bosch-Ramon}, V. 2019, \aap, 623, A91

\bibitem[{{van Dokkum} \& {Conroy}(2010)}]{Dokkum_2010}
{van Dokkum}, P.~G. \& {Conroy}, C. 2010, \nat, 468, 940

\bibitem[{{Vieyro} {et~al.}(2017){Vieyro}, {Torres-Alb{\`a}}, \&
  {Bosch-Ramon}}]{Vieyro_2017}
{Vieyro}, F.~L., {Torres-Alb{\`a}}, N., \& {Bosch-Ramon}, V. 2017, \aap, 604,
  A57

\bibitem[{{Vlahakis} \& {K{\"o}nigl}(2004)}]{Vlahakis_2004}
{Vlahakis}, N. \& {K{\"o}nigl}, A. 2004, \apj, 605, 656

\bibitem[{{Wang} {et~al.}(2022){Wang}, {Liu}, {Petropoulou}, {Oikonomou},
  {Xue}, \& {Wang}}]{Wang_2022}
{Wang}, Z.-R., {Liu}, R.-Y., {Petropoulou}, M., {et~al.} 2022, \prd, 105,
  023005

\bibitem[{{Wykes} {et~al.}(2013){Wykes}, {Croston}, {Hardcastle}, {Eilek},
  {Biermann}, {Achterberg}, {Bray}, {Lazarian}, {Haverkorn}, {Protheroe}, \&
  {Bromberg}}]{Wykes_2013}
{Wykes}, S., {Croston}, J.~H., {Hardcastle}, M.~J., {et~al.} 2013, \aap, 558,
  A19

\bibitem[{Wykes {et~al.}(2014)Wykes, Hardcastle, Karakas, \& Vink}]{Wykes_2014}
Wykes, S., Hardcastle, M.~J., Karakas, A.~I., \& Vink, J.~S. 2014, MNRAS, 447,
  1001

\bibitem[{{Wykes} {et~al.}(2018){Wykes}, {Taylor}, {Bray}, {Hardcastle}, \&
  {Hillas}}]{Wykes_2018}
{Wykes}, S., {Taylor}, A.~M., {Bray}, J.~D., {Hardcastle}, M.~J., \& {Hillas},
  M. 2018, Nuclear and Particle Physics Proceedings, 297-299, 234

\bibitem[{{Zhu} {et~al.}(2010){Zhu}, {Blanton}, \& {Moustakas}}]{Zhu_2010}
{Zhu}, G., {Blanton}, M.~R., \& {Moustakas}, J. 2010, \apj, 722, 491

\end{thebibliography}

\end{document}